%% file: diving.tex
\documentclass{elsart}
\usepackage{amsmath,amssymb,eucal,mathrsfs}
\usepackage[dvips]{graphicx}
\usepackage{lscape,palatino,caption,listings,url,natbib,algorithmic,multirow,longtable,fancyvrb} 
\usepackage[%
  pdfauthor={Jakub Marecek},%
  pdftitle={Decomposition, Reformulation, and Diving in University Course Timetabling},%
  pdfkeywords={Integer programming, Soft Constraints, Diving, Heuristic, Hybrid, Metaheuristic, Timetabling, University Course Timetabling, Udine},%
  pdfstartview=FitH,%
  bookmarks=true,%
  bookmarksopen=true,%
  breaklinks=true,%
  colorlinks=true,%
  linkcolor=blue,anchorcolor=blue,%
  citecolor=blue,filecolor=blue,%
  menucolor=blue,pagecolor=blue,%
  urlcolor=blue]{hyperref}

\include{commands}

\begin{document}

\begin{frontmatter}
\title{Decomposition, Reformulation, and Diving \\ in University Course Timetabling}
\author{Edmund K. Burke},
\author{Jakub Mare{\v c}ek},
\author{Andrew J. Parkes}
\address{School of Computer Science, University of Nottingham \\ Jubilee Campus, Nottingham NG8 1BB, UK} 
\author{Hana Rudov{\' a}}
\address{Masaryk University Faculty of Informatics \\ Botanick{\' a} 68a, Brno 602 00, The Czech Republic}

\begin{abstract}
In many real-life optimisation problems, there are multiple interacting components in a solution.
For example, different components might specify assignments to different kinds of resource. 
Often, each component is associated with different sets of soft constraints, and so with 
different measures of soft constraint violation. The goal is then to minimise a linear combination of such
measures. 
This paper studies an approach to such problems, which can be thought of as multiphase 
 exploitation of multiple objective-/value-restricted submodels. 
In this approach, only one computationally difficult component of a problem and the associated subset 
 of objectives is considered at first.
This produces partial solutions, 
 which define interesting neighbourhoods in the search space of the complete problem.
Often, it is possible to pick the initial component so that 
 variable aggregation can be performed at the first stage, 
 and the neighbourhoods to be explored next are guaranteed to contain feasible solutions. 
Using integer programming, it is then easy 
 to implement heuristics producing solutions with bounds on their quality.

Our study is performed on a university course timetabling problem used in
 the 2007 International Timetabling Competition,
 also known as the Udine Course Timetabling Problem. 
The goal is to find an assignment of events to periods and rooms, so that 
 the assignment of events to periods is a feasible bounded colouring of an associated
 conflict graph and the linear combination of
 the numbers of violations of four soft constraints is minimised.
In the proposed heuristic,
 an objective-restricted neighbourhood generator produces assignments of periods to events,
 with decreasing numbers of violations of 
 two period-related soft constraints.
Those are relaxed into assignments of events to days, 
 which define neighbourhoods that are easier to search with respect 
 to all four soft constraints.
Integer programming formulations for all subproblems are given
  and evaluated using ILOG CPLEX 11.
The wider applicability of this approach is analysed and discussed. 
\end{abstract}

\begin{keyword}
Integer programming \sep Decomposition \sep Reformulation \sep Diving \sep Heuristic \sep Metaheuristic \sep University Course Timetabling \sep Soft Constraints 
\end{keyword}

\end{frontmatter}

\input{intro}
\input{problem}

\input{ip-monolithic}

\input{overview}

\input{overview2}
\input{ip-subproblems}

\input{related}

\input{results}

\small{
\bibliography{diving}
}

\end{document}

%% file: commands.tex
\long\def\IGNORE#1{}
\long\def\TODO#1{}
\long\def\NOTE#1{}

\DeclareMathOperator{\Weights}{W}
\DeclareMathOperator{\WCapacity}{W^{RCap}}
\DeclareMathOperator{\WSpread}{W^{TSpr}}
\DeclareMathOperator{\WCompactness}{W^{TCom}}
\DeclareMathOperator{\WStability}{W^{RStb}}
\DeclareMathOperator{\PCapacity}{P^{RCap}}
\DeclareMathOperator{\PSpread}{P^{TSpr}}
\DeclareMathOperator{\PCompactness}{P^{TCom}}
\DeclareMathOperator{\PStability}{P^{RStb}}

\DeclareMathOperator{\Rooms}{R}
\DeclareMathOperator{\MRooms}{M}
\newcommand{\HasCapacity}[1]{\textnormal{A}^{#1}}

\DeclareMathOperator{\Courses}{C}
\newcommand{\HasMinDays}[1]{\textnormal{Q}^{#1}}

\newcommand{\HasEventsCount}[1]{\textnormal{E}^{#1}}
\newcommand{\HasStudents}[1]{\textnormal{S}^{#1}}

\DeclareMathOperator{\Curricula}{U}
\newcommand{\CurriculumHasCourses}[1]{\textnormal{\ensuremath{\tilde  U}}^{#1}}

\DeclareMathOperator{\Periods}{P}
\DeclareMathOperator{\Days}{D}
\newcommand{\HasPeriods}[1]{\textnormal{\ensuremath{\tilde D}}^{#1}}

\DeclareMathOperator{\SingletonChecks}{Z}

\DeclareMathOperator{\Teachers}{T}
\newcommand{\TeacherHasCourses}[1]{\textnormal{\ensuremath{\tilde T}}^{#1}}

\DeclareMathOperator{\Deprecated}{N}
\newcommand{\Taught}[1]{x_{#1}}

\newcommand{\CourseMinDaysViolations}[1]{q_{#1}}
\newcommand{\Singletons}[1]{z_{#1}}
\newcommand{\SetTimes}[1]{w_{#1}}
\newcommand{\SetDays}[1]{D^{#1}}
\newcommand{\CourseSchedule}[1]{v_{#1}}
\newcommand{\CourseRooms}[1]{y_{#1}}

\newcommand{\abs}[1]{\left| {#1} \right|}

%% file: intro.tex
\long\def\ignore#1{}
\long\def\NOTE#1{{\it{[ AJP: #1]}}}
\long\def\PARA#1{{\bf{ ..... #1 ....}}}
\let\NOTE=\ignore  \let\PARA=\ignore

\section{Introduction}

 \NOTE{ I have added lots of things that I think help to explain the
 background of the overall approach, so it is still quite disjointed,
 and probably needs re-ordering. However, we should see if we agree on
 the overall content now, before trying to fix this further.  }

\PARA{General ideas of reduction and large neighbourhoods}

\subsection{Motivation}

There has recently been considerable progress in the development of
mixed integer programming solvers
\citep{NemhauserWolsey1988,Nemhauser2000,Bixby2004}. To a considerable
extent, this progress is due to the introduction of new primal and
improvement heuristics
\citep{Fischetti2003,Fischetti2005,Danna2005,Berthold2007}.  
These methods can be said to ``dive'' into a particular region of the
search space in order to explore it intensely. This is intended to
restrict and simplify the search space so that the sub-instance is
easier to solve. 
Currently, diving heuristics are run only once the initial linear
programming relaxation is solved: Polynomial-time reduction strategies 
\citep{Lin1965}
use the relaxed solution to identify a subset of variables whose
values should be fixed.
Typically, and inevitably, the dive will remove the optimal solutions;
the optimum of the reduced instance is not necessarily optimal in the
original instance.  It is hoped, nevertheless, that improving integer
feasible solutions can still be extracted from solutions of the
modified instance.
Despite substantial recent progress, we believe that there is
still considerable scope for further work in this general direction.

The approach we propose is based on the observation that there is a complex structure 
to many real-world problems. The objective function is usually highly structured in
itself, and hence not best thought of as the generic $c^T x$ form, which is the usual starting
point of theoretical analyses.  In many instances, the overall objective function is a linear
combination of many disparate terms with each term representing a
total penalty for some unwanted structure within a possibly small part of a solution,
 or profit from some desirable structure.
Furthermore, the different terms in the objective are rarely equally hard to
handle. Some can be relatively easy, but some terms can be
troublesome; representing them can require a large number of auxiliary
variables. For example, some term might well contain a penalty that 
uses a ``sum of absolute values'' or ``max of a max'' structure that can require many extra
variables to compute. 
Hence, the first ingredient of our approach is to 
generate sub-instances by selectively keeping some terms in the objective. 
Which terms should be kept in the objective, will become clearer once we sketch
out the rest of our approach.  

The second ingredient is the aggregation of variables: Defining a new variable to be some
linear or simple non-linear function of existing variables, such as the maximum
of a small subset. In many cases, a human expert inspecting the sub-instance would suggest 
that particular aggregations of variables would allow it to be rewritten into a form
that is much easier to solve. 
Current diving heuristics, however, do not perform any aggregations that are specific to the diving,
despite the potential to greatly reduce the number of variables and, furthermore, to do so in a
manner that is quite different than reductions by means of fixing values of variables.
We distinguish different potential intended usages of aggregation as
follows:
\begin{itemize}

 \item \textbf{Equivalence-based}. The new variable is 'logically equivalent to
 the old variables' in the sense that it can be used to replace them,
 and still preserve optimality and validity.  This is the type of
 aggregation that is performed in preprocessing.

 \item \textbf{Validity-based}. The new aggregate variable is not
 intended to replace the old variables fully: They either supplement the 
 original ones, or can be used to provide valid cuts and valid lower bounds. 

 \item \textbf{Solution-based}.  The new aggregate variable is not
 intended to replace the old variables fully: 
 They either supplement the original ones, or can be used to provide
 valid solutions (upper bounds). This, in a certain sense, can be thought of as 
 the dual of validity-based aggregation.

\end{itemize}
This concept of validity-based aggregation underlies many of the constructions in this paper.

Finally, current diving heuristics seem to have linear programming relaxation as
the input and (possibly) an integer feasible solution as the output.  
We suggest a different approach, (possibly) producing integer feasible solutions
together with lower bounds for the original instance, without utilising its linear 
programming relaxation.  
This makes it possible to run the diving heuristic prior
to computing the first linear programming relaxation of the original instance 
and improve the performance of subsequent cut generation. 
In step one of our approach, a much smaller (``surface'') instance of integer
programming is derived by picking suitable components of the original instance
and applying the validity-based aggregation introduced above.
By solving this instance, we obtain a valid lower bound for the original instance,
in addition to a feasible input to the second step.
In step two, we dive into a neighbourhood obtained by reducing the feasible solution
of step one. 
Notice that the (``dive'') neighbourhood can often be much smaller compared to the original
search space, whilst still being guaranteed to contain a feasible solution to the
original instance.
Various strategies can be used to control the order of performing the individual dives.
In the long run, it would make sense to explore the possibilities of
applying various strategies to initialise and perform dives, 
automating the choice of an aggregating reduction strategy,
as well as of tailoring automated reformulations better to instances obtained in such
reduction strategies. 

\begin{figure}[!t]
\centering
\begin{tabular}{|l|c|c|c|}
\hline
Offers / Model & Monolithic & Surface & Dive \\
\hline
Lower bounds & $\checkmark$          & $\checkmark$ &        \\
Upper bounds & $\checkmark$          &           & $\checkmark$    \\
Completeness & $\checkmark$          &           & \\
\hline 
\end{tabular}
\caption{A schematic overview of the suggested decomposition: As a primal heuristic for the convergent 
monolithic model, we employ an aggregated surface model and a series of dives, bounded by the solution
obtained at surface.}
\vskip 6mm
\label{tab:intro}
\end{figure}

\subsection{The Result}

In this paper, we make an initial step in this direction,
giving a ``proof of concept'' implementation for a particular class of
instances from timetabling.  
In step one of our approach, a much smaller instance of integer
programming is solved.
Feasible solutions of the smaller instance define the neighbourhoods
to dive into in step two.
The smaller instance is derived using the ``validity-based aggregation'',
introduced above.
We also hint at the power of
so-far-non-automated reformulations on instances obtained using such
reduction strategies.

In particular, we propose such a heuristic decomposition for Udine Course Timetabling,
 a benchmark course timetabling problem,
 used in Track 3 of the International Timetabling Competition 2007 \citep{DiGaspero2006}.
There, as in many other timetabling problems,
 feasible solutions are determined by a bounded graph colouring component \citep{Welsh1967,Burke2004}, 
 more specifically by an extension of a pre-colouring bounded in the number of uses of a colour, 
 which is difficult in its own right \citep{Even1976,Garey1979,Bodlaender2005}.
The quality of feasible solutions is measured by the number of violations 
  of four additional complex soft constraints.
These soft constraints place emphasis on:
\begin{itemize}
 \item suitability of rooms with respect to their capacities
 \item suitability of the spread of the events of a course within the weekly timetable
 \item minimisation of the number of distinct rooms each course uses
 \item desirability of various patterns in distinct individual daily timetables.
\end{itemize} 
Note that the latter three soft constraints will be formulated using the $\max$ function. 
Modelling of the soft constraints, as opposed to the bounded graph colouring component on its own,
 entails an increase in the dimensions of the model, and consequently of the run-time, making 
 even modest real-life instances difficult to solve using a stand-alone integer programming solver 
 \citep{Marecek2007GOR,Marecek2007TR},
 or branch-and-cut procedures \citep{Avella2005,Marecek2008TR}.

Hence, we propose a heuristic that decomposes the process of solving the instance into two stages, 
 which are outlined in Table~\ref{tab:intro}. 
The ``surface'' stage 
 is restricted to the bounded colouring problem with the two period-related soft constraints. 
The ``dives'' 
 fix certain features of the solution found at the surface,
 and produce a locally optimal solution to the complete instance. 
We employ two ``control strategies'' for executing the dives. 
In ``anytime strategies'', dives are executed whenever a feasible bounded colouring is found at the surface.
In ``contract strategies'', a time limit for the search at the surface is given, and dives are 
 executed only afterwards.
As it seems, previously underutilised ``contract strategies'' with multiple 
  ``neighbourhoods'' searched in the dives in a suitable order
  perform considerably better than the traditional anytime strategies 
  using a single type of restriction. 
 
A solver based on this approach was implemented using ILOG Concert and ILOG CPLEX 11.
Using non-automatically reduced integer programming formulations, 
 it yields good solutions to instances of up to 434 events and 81 distinct enrolments (``curricula'') 
 within thirty minutes of run time on a desktop PC.
The lower bounds obtained at the surface 
 are better than those which \cite{Luebbecke2008} obtain within the same time limits.
Overall, the heuristic produces solutions for the Udine Course Timetabling Problem,
 together with bounds on their distance from optimality. 
 
The two-fold aims of this paper are reflected in its structure. 
The immediate goal is to obtain better solutions and lower bounds to the Udine Course Timetabling problem.
The problem is introduced in Section~\ref{sec:problem}.
The particular heuristic is outlined in Section~\ref{sec:over2}.
The integer programming formulations of the subproblems we employ are presented in Section~\ref{sec:subproblems}.
Finally, computational experience is described in Section~\ref{sec:results}.
We regard this as part of a longer-term goal, however, which is the development of methods
  to better control and exploit various, possibly automated, decompositions and reformulations. 
The methodology of ``Multiphase Exploitation of Multiple Objective-/value-restricted Submodels'' 
  (or ``MEMOS'' for short) is outlined in Section~\ref{sec:over}.
We are not claiming that any of the individual techniques presented in Section~\ref{sec:over} are novel {\it per se}. 
Rather, we are providing a basis for a rationalisation and classification of previously known methods,
  which are surveyed in Section~\ref{sec:related}. 
Both the particular solver and the more generally applicable approach 
 to the design of hybrid metaheuristics for complex problems may be of separate interest.


%% file: problem.tex
\section{Problem Description}
\label{sec:problem}

\subsection{University Course Timetabling}

In general, timetabling problems share the search for feasible colourings of
 conflict graphs, where vertices represent events and there is an edge between
 two vertices if the corresponding events cannot take place at the same time
 \citep{Burke2004}. 
This graph colouring component is often bounded in the number of uses of a colour
 and has to provide an extension of pre-coloured conflict graph, such that the 
 total number of violations of certain soft constraints is minimised. 
In university course timetabling \citep{Bardadym1996,Burke1997,Petrovic2004,McCollum2006}, 
 these soft constraints often stipulate that events should be timetabled for rooms 
 of appropriate sizes.
At least or at most a certain number of days of instructions 
 should be timetabled for groups of students and individual teachers,
 and daily timetables of students or teachers should not exhibit particular patterns.
For example, a single event per day or long gaps in a daily timetable, or on the other 
 hand, six events per day with no gap around lunch time may be deemed undesirable. 
The particulars of each problem instance vary widely from university to university, 
 and a number of both exact and heuristic search methods have been 
 attempted on a range of instances. For surveys, see \citet{Carter1997,Schaerf1999,Petrovic2002}.
Out of the numerous variations of the problem in use, instances from
 the University of Udine \citep{DiGaspero2003}
 and Purdue University \citep{Rudova2002,Rudova2007},
 together with random problem instances used in the International Timetabling 
 Competition\footnote{See \citet{DiGaspero2007TR} or \url{http://www.cs.qub.ac.uk/itc2007/} for International Timetabling Competition (ITC) 2007 and \url{http://www.idsia.ch/Files/ttcomp2002/} for ITC 2002.} in 2002 and 2007  
 have recently started to be considered as benchmark problems.

\begin{figure}[!t]
\centering
\includegraphics[height=0.4\textheight]{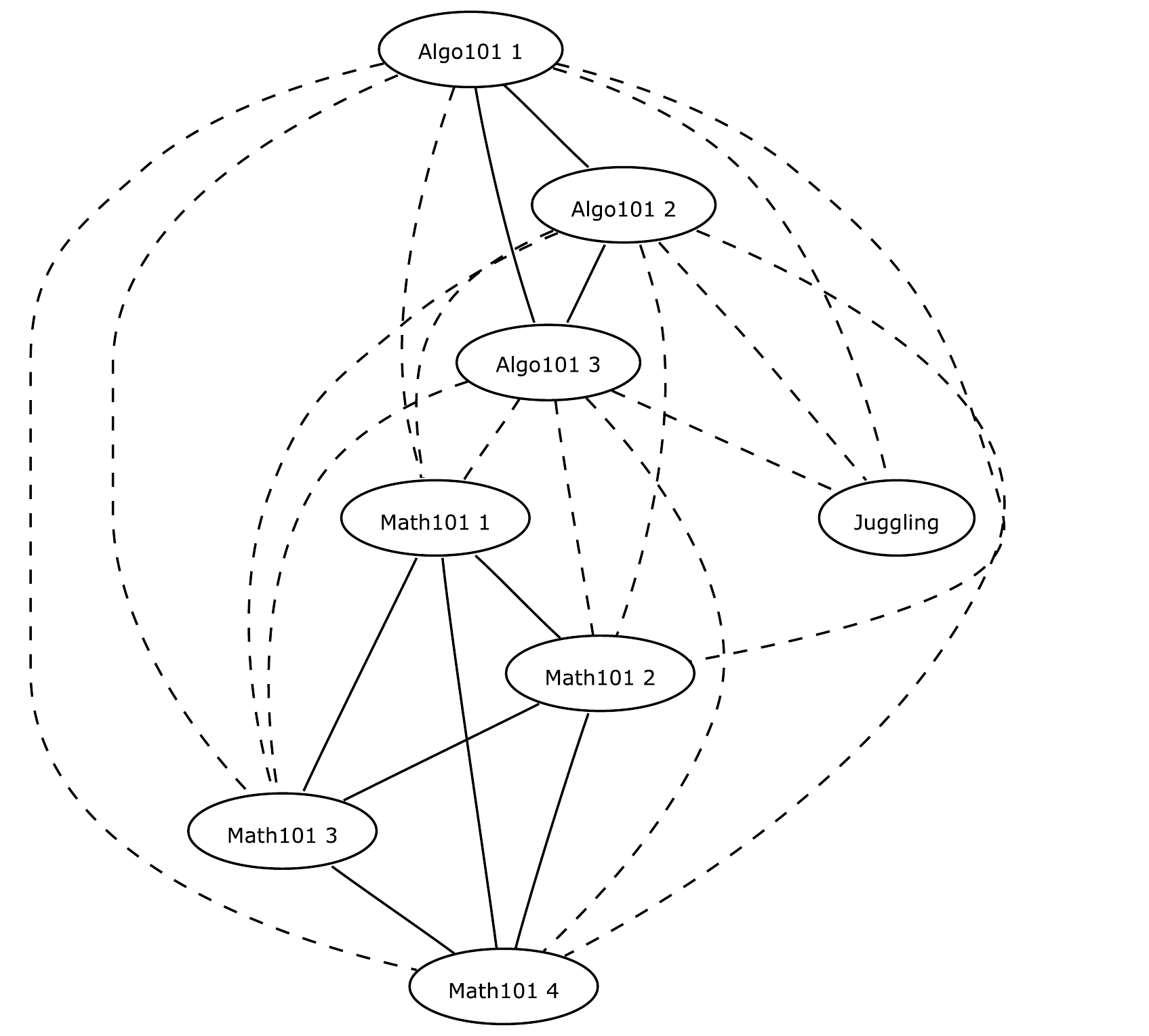}
\caption{
The colouring component of a trivial instance of course timetabling:
There are three courses, Juggling, Math101 and Algo101, with 1, 4 and 3 events respectively.
One group of students attends both Math101 and Algo101,  
 another group attends both Juggling and Algo101,
 and events attended by a group of students cannot overlap in time.
Each vertex in the associated conflict graph represents an event and there is an edge between two vertices, 
 if the two corresponding events cannot take place concurrently. 
The edge is dashed, if this requirement is given by the enrolment of a student in both courses.}
\label{fig:timetabling}
\vskip 6mm 
\end{figure} 
 
\subsection{The Udine Course Timetabling Problem}
The Udine Course Timetabling problem studied in this paper
is maintained by \citet{DiGaspero2003} at \emph{Universit{\` a} degli studi di Udine}.
There are two important assumptions:
\begin{itemize}
  \item Events are partitioned into disjoint subsets, called courses; 
  events of any one course have to take place at different times,
  are attended by the same number of students,  
  and are freely interchangeable 
  \item A small number of distinct, possibly overlapping, sets of courses,
  representing enrolments prescribed to various groups of students, 
  are identified and referred to as curricula.
\end{itemize}
Due to the second assumption, this problem is often referred to as ``curriculum
based timetabling'', as opposed to ``student enrolment based timetabling'', which
tries to minimise the number of conflicts among a possibly large number of enrolments.
See Figure~\ref{fig:timetabling} for an illustrative example. 
The complete input can be captured by seven constant sets and eight mappings:
\begin{itemize}
\item $\Courses$, $\Curricula$, $\Teachers$, $\Rooms$, $\Days$, $\Periods$ are sets of courses, curricula,
      teachers, rooms, days, and periods, respectively
\item $\CurriculumHasCourses{u}$ is the non-empty set of courses in curriculum $u$ 
\item $\Deprecated$ is a subset of $\Courses \times \Periods$, giving forbidden course-period combinations
\item $\HasEventsCount{c}$ is the number of events course $c$ has in a week
\item $\HasStudents{c}$ is the number of students enrolled in course $c$ 
\item $\HasMinDays{c}$ is the prescribed minimum number of distinct week-days of instruction for course $c$   
\item $\TeacherHasCourses{t}$ is the subset of courses $\Courses$ taught by teacher $t$
\item $\HasCapacity{r}$ is the capacity of room $r$ 
\item $\HasPeriods{d}$ is the subtuple (ordered subset) of $\Periods$ corresponding to periods in day $d$ 
\item $\Weights$ is a vector of non-negative weights $(\WCapacity, \WSpread, \WCompactness, \WStability)$ for the four soft constraints described below.
\end{itemize}

\begin{table}[!b]
\centering
\renewcommand{\arraystretch}{1.2}
\caption{
Instances of the Udine Course Timetabling problem: numbers of rooms and periods; 
the number of courses and the sum of their events in a week;
frequency, or the portion of period-room slots in use, and utilisation, or
the portion of period-seat slots in use \citep{Beyrouthy2007};
the number of distinct enrolments (``curricula'');
numbers of edges and density in conflict graphs (CG) with vertices representing courses, rather than events \citep{Marecek2007TR}.
}
\label{tab:instances}
\vskip 6mm
\input{tables/instances}

\end{table}

Informally, the goal is to produce a mapping from events to period-rooms pairs such that:
\begin{enumerate} 
\item For each course $c$, $\HasEventsCount{c}$ events are timetabled \label{itm:hard1} 
\item No two events take place in the same room in the same period
\item No two events of a single course, no two events taught by a single teacher, and no two events included in a single curriculum are taught at the same time \label{itm:hard3}
\item No event of course $c$ is taught in a period $p$, if  $\left\langle {c, p} \right\rangle$ is in $\Deprecated$ \label{itm:inconvenient}
\item The objective is to minimise a weighted sum of penalty terms $$\WCapacity \PCapacity + \PSpread \WSpread + \PCompactness \WCompactness + \PStability \WStability,$$ where: 
\begin{itemize}
\item $\PCapacity$ (for ``room capacity'') is the number of students left without a seat at an event, 
      summed across all events; this is the value of the number of students attending an event
      minus the capacity of the allocated room over all events where the value is positive \label{itm:overspill}
\item $\PSpread$ (for ``spread of events of a course over distinct week-days'') 
      sums the value of the number of prescribed distinct week-days of instruction minus 
      the actual number of distinct week-days of instruction over all courses where the value is positive \label{itm:violations} 
\item $\PCompactness$ (for ``time compactness'') is the number of isolated events  
      in daily timetables of individual curricula; 
      ``[f]or a given curriculum we account for a violation every time there 
      is one lecture not adjacent to any other lecture on the same day'' \citep{DiGaspero2007TR} \label{itm:patterns}
\item $\PStability$ (for ``room stability'') is the number of distinct course-room allocations 
       on the top of a single course-room allocation per course. \label{itm:stability}      
\end{itemize}
\end{enumerate}

In the original paper of \citet{DiGaspero2003}, there were described only four instances 
 of the Udine Course Timetabling Problem of up to 252 events and 57 distinct enrolments,
 with $(\WCapacity, \WSpread, \WCompactness, \WStability) = (1, 5, 2, 0)$.
Fourteen more instances of up to 434 events and 81 distinct enrolments have now 
 been made available in Track 3 of the International Timetabling Competition,
 with weights $(\WCapacity, \WSpread, \WCompactness, \WStability) = (1, 5, 2, 1)$. 
Their dimensions are summarised in Table~\ref{tab:instances}. 

\NOTE{AJP: is it standard in this journal to put table captions above table? It sometimes gives weird formatting.}

%% file: tables/instances.tex
\begin{tabular}{cc|ccccrrccr}
Instance & AKA & \rotatebox{90}{Rooms} & \rotatebox{90}{Periods} & \rotatebox{90}{Courses} & \rotatebox{90}{Events} & \rotatebox{90}{Frequency} \rotatebox{90}{(used slots)} & \rotatebox{90}{Utilisation} \rotatebox{90}{(used seats)} & \rotatebox{90}{Curricula} & \rotatebox{90}{Edges in CG} \rotatebox{90}{(course-based)} & \rotatebox{90}{Density of CG} \rotatebox{90}{(course-based)} \\ \hline
comp01 & Fis0506-1 & 6 & 30 & 30 & 160 & 88.89 \% & 45.98 \% & 14 & 53 & 12.18 \% \\
comp02 & Ing0203-2 & 16 & 25 & 82 & 283 & 70.75 \% & 46.28 \% & 70 & 401 & 12.07 \% \\
comp03 & Ing0304-1 & 16 & 25 & 72 & 251 & 62.75 \% & 38.30 \% & 68 & 342 & 13.38 \% \\
comp04 & Ing0405-3 & 18 & 25 & 79 & 286 & 63.56 \% & 33.22 \% & 57 & 212 & 6.88 \% \\
comp05 & Let0405-1 & 9 & 36 & 54 & 152 & 46.91 \% & 43.50 \% & 139 & 917 & 64.08 \% \\
comp06 & Ing0506-1 & 18 & 25 & 108 & 361 & 80.22 \% & 45.28 \% & 70 & 437 & 7.56 \% \\
comp07 & Ing0607-2 & 20 & 25 & 131 & 434 & 86.80 \% & 41.71 \% & 77 & 508 & 5.97 \% \\
comp08 & Ing0607-3 & 18 & 25 & 86 & 324 & 72.00 \% & 37.39 \% & 61 & 214 & 5.85 \% \\
comp09 & Ing0304-3 & 18 & 25 & 76 & 279 & 62.00 \% & 32.67 \% & 75 & 251 & 8.81 \% \\
comp10 & Ing0405-2 & 18 & 25 & 115 & 370 & 82.22 \% & 36.38 \% & 67 & 481 & 7.34 \% \\
comp11 & Fis0506-2 & 5 & 45 & 30 & 162 & 72.00 \% & 56.23 \% & 13 & 75 & 17.24 \% \\
comp12 & Let0506-2 & 11 & 36 & 88 & 218 & 55.05 \% & 35.06 \% & 150 & 1181 & 30.85 \% \\
comp13 & Ing0506-3 & 19 & 25 & 82 & 308 & 64.84 \% & 38.14 \% & 66 & 216 & 6.50 \% \\
comp14 & Ing0708-1 & 17 & 25 & 85 & 275 & 64.71 \% & 34.78 \% & 60 & 368 & 10.31 \% \\
\end{tabular}

%% file: ip-monolithic.tex
\subsection{An Integer Programming Formulation}
\label{sec:monolithic}

Formally, the Udine Course Timetabling problem can be described 
  using an integer programming model. 
Doing so necessitates the choice of decision variables. 
This is of little importance if the sole purpose is to formulate the problem formally,
  but becomes of paramount importance, if the performance of the formulation is evaluated.
It seems tempting to see the problem as a variation of the three-index assignment 
 and to use binary variables for each event-room-period combination.
This corresponds to the trivial formulation of graph colouring, where the number of binary variables  
 is the product of the number of vertices and an upper bound on the number of colours.
There are, however, many alternative formulations of graph colouring \citep{Marecek2007TR} 
 and it seems reasonable to use the best available formulation of graph colouring
 that would admit formulation of the soft constraints. 
After exploring a number of such alternatives, \citet{Marecek2007TR} proposed
 a formulation based on a suitable clique-partition, which is given implicitly in many 
 graph colouring applications.
For Udine Course Timetabling, this formulation translates to a smaller number
 of ``core'' binary decision variables $\Taught{}$, given by the product of the numbers periods, rooms,
 and courses, rather than events.
Outside of those, there are dependent variables $\CourseSchedule{}$, $\CourseMinDaysViolations{}$, 
 $\Singletons{}$, and $\CourseRooms{}$, whose values are derived from the 
 values of $\Taught{}$ in the process of solving:
\TODO{AJP wants to rewrite this. JXM doesn't.}
\begin{itemize} 
\item $\Taught{}$ are binary decision variables indexed with periods, rooms and courses. 
Their values are constrained so that in any feasible solution, course $c$ should be taught in room $r$ at period $p$,
 if and only if $\Taught{p,r,c}$ is set to one.
\item $\CourseSchedule{}$ are binary decision variables indexed with courses and days.
Their values are constrained so that in any feasible solution, there is at least one event of course $c$ held on day $d$,
  if and only if $\CourseSchedule{d,c}$ is set to one.
\item $\CourseMinDaysViolations{}$ are integer decision variables indexed with courses,
 whose values are bounded below by zero and above by the number of days in a week.  
Their values are constrained so that in any feasible solution, $\CourseMinDaysViolations{c}$ is the 
 number of days course $c$ is short of the recommended 
 days of instruction, $\HasMinDays{c}$.
\item $\Singletons{}$ are binary decision variables indexed with curricula,
 days, and an index-set $\SingletonChecks$ of natural pattern-penalising constraints. 
Their values are constrained so that in any feasible solution, $\Singletons{u,d,s}$ is set to one
 if and only if the pattern-penalising constraint indexed by $s \in \SingletonChecks$ 
 is violated in the timetable for curriculum $u$ and day $d$ given by the solution.
\item $\CourseRooms{}$ are binary decision variables indexed with rooms and courses.
Their values are constrained so that in any feasible solution, $\CourseRooms{r,c}$ is set to one
 if and only if room $r$ is used by course $c$.       
\end{itemize}

The objective function can be expressed as:
\begin{align}  
  \min & \; \WCapacity \sum_{r \in \Rooms} \sum_{p \in \Periods} \sum_{\substack{c \in \Courses \\ \HasStudents{c} > \HasCapacity{r}}} 
       \Taught{p,r,c} \; (\HasStudents{c} - \HasCapacity{r})  
     + \; \WCompactness \sum_{u \in \Curricula} \sum_{d \in \Days} \sum_{s \in \SingletonChecks} \Singletons{u,d,s} \notag \\ 
     + & \; \WSpread \sum_{c \in \Courses} \CourseMinDaysViolations{c} 
     + \; \WStability \sum_{c \in \Courses} \left( {\left( {\sum_{r \in \Rooms} \CourseRooms{r,c}} \right) - 1} \right) \notag
\end{align}

Hard constraints can be formulated as follows:
\begin{align}
  \label{eqn:hardFirst}
  \hskip 2cm \forall c \in \Courses \hskip 2cm & \sum_{p \in \Periods} \sum_{r \in \Rooms} & \Taught{p,r,c} & = \HasEventsCount{c} \\
  \label{eqn:hardSecond}
  \hskip 2cm \forall p \in \Periods \forall r \in \Rooms \hskip 2cm & \sum_{c \in \Courses} & \Taught{p,r,c} & \le 1 \\  
  \label{eqn:hardThird}
  \hskip 2cm \forall p \in \Periods \forall c \in \Courses \hskip 2cm & \sum_{r \in \Rooms} & \Taught{p,r,c} & \le 1 \\
  \label{eqn:hardFourth}
  \hskip 2cm \forall p \in \Periods \forall t \in \Teachers \hskip 2cm & \sum_{r \in \Rooms} \sum_{c \in \TeacherHasCourses{t}} & \Taught{p,r,c} & \le 1 \\ 
  \label{eqn:hardFifth}
  \hskip 2cm \forall p \in \Periods \forall u \in \Curricula \hskip 2cm & \sum_{r \in \Rooms} \sum_{c \in \CurriculumHasCourses{u}} & \Taught{p,r,c} & \le 1 \\
  \label{eqn:hardLast}
  \hskip 2cm \forall \left\langle {c, p} \right\rangle \in \Deprecated \hskip 2cm & \sum_{r \in \Rooms} & \Taught{p,r,c} & = 0  
\end{align}
Constraint (\ref{eqn:hardFirst}) enforces a given number of events to be taught for each course.
Constraint (\ref{eqn:hardSecond}) ensures no two events are taught in a single room at a single period.
Constraints (\ref{eqn:hardThird}--\ref{eqn:hardFifth}) stipulate that only one 
 event within a single course or curriculum, or taught by a single teacher can be held at any given period.
Notice the similarity of constraints (\ref{eqn:hardFirst}--\ref{eqn:hardFifth}) and
 the clique-based formulation of bounded graph colouring \citep{Marecek2007TR}. 
Finally, constraint (\ref{eqn:hardLast}) forbids the use of some periods in timetables 
 of some courses, corresponding to a pre-colouring extension.  
Notice also that constraint (\ref{eqn:hardFifth}) renders constraint (\ref{eqn:hardThird}) redundant, 
 if there are no courses outside of any curricula.

The formulation of soft constraints is less trivial and, as will be shown in Section~\ref{sec:results} 
 and Table \ref{tab:lp-relaxations}, the formulation presented below proves to be quite
 challenging for modern general purpose integer programming solvers, even after the 
 strengthening proposed by \citet{Marecek2008TR}. 
Values of $\CourseSchedule{}$ are forced to maxima of certain subsets of $\Taught{}$ 
 using constraints (\ref{eqn:courseshed1}--\ref{eqn:courseshed2}),
 in effect constructing daily timetables for individual curricula.
Subsequently, the number of distinct week-days of instruction that course $c$ is short of the 
 prescribed value $\HasMinDays{c}$, can be forced into $\CourseMinDaysViolations{c}$ using
 constraint (\ref{eqn:courseshed3}):
\begin{align}
  \label{eqn:courseshed1}
  \hskip 2cm \forall c \in \Courses \forall d \in \Days \forall p \in \HasPeriods{d} &
  \hskip 2cm \sum_{r \in \Rooms}
  \Taught{p,r,c} & \le \CourseSchedule{d,c} \\  
  \label{eqn:courseshed2}
  \hskip 2cm \forall c \in \Courses \forall d \in \Days &
  \hskip 2cm \sum_{r \in \Rooms} \sum_{p \in \HasPeriods{d}}
  \Taught{p,r,c} & \ge \CourseSchedule{d,c} \\
  \label{eqn:courseshed3}
  \hskip 2cm \forall c \in \Courses &
  \hskip 2cm \sum_{d \in \Days} \CourseSchedule{d,c} & \ge \HasMinDays{c} - \CourseMinDaysViolations{c}
\end{align}

\TODO{AJP: Needs an explanation that we use patterns in CUTS.}
The ``natural'' formulation of penalisation of patterns \citep{Marecek2007GOR} 
 occurring in daily timetables of individual curricula goes through the daily 
 timetables bit by bit, first by checking isolated events in the first and the last 
 period of the day, and later looking for triples of consecutive periods with
 only the middle period occupied by an event:
For an instance with four periods per day, this is:
\begin{align}
  \label{eqn:patmat1}
  \forall u \in \Curricula, d \in \Days, \forall \left\langle {p_1, p_2, p_3, p_4} \right\rangle \in \HasPeriods{d} &
  \; \sum_{c \in \CurriculumHasCourses{u}} \sum_{r \in \Rooms} &
  (\Taught{p_1,r,c} - \Taught{p_2,r,c}) & \le \Singletons{u,d,1}
\\
  \label{eqn:patmat2}
  \forall u \in \Curricula, d \in \Days, \forall \left\langle {p_1, p_2, p_3, p_4} \right\rangle \in \HasPeriods{d} &    
  \; \sum_{c \in \CurriculumHasCourses{u}} \sum_{r \in \Rooms} &  
  (\Taught{p_4,r,c} - \Taught{p_3,r,c}) & \le \Singletons{u,d,2}
\\
  \label{eqn:patmat3}
  \forall u \in \Curricula, d \in \Days, \forall \left\langle {p_1, p_2, p_3, p_4} \right\rangle \in \HasPeriods{d} &    
  \; \sum_{c \in \CurriculumHasCourses{u}} \sum_{r \in \Rooms} & 
  (\Taught{p_2,r,c} - \Taught{p_1,r,c} - \Taught{p_3,r,c}) & \le \Singletons{u,d,3}
\\
  \label{eqn:patmat4}
  \forall u \in \Curricula, d \in \Days, \forall \left\langle {p_1, p_2, p_3, p_4} \right\rangle \in \HasPeriods{d} & 
  \; \sum_{c \in \CurriculumHasCourses{u}} \sum_{r \in \Rooms} & 
  (\Taught{p_3,r,c} - \Taught{p_2,r,c} - \Taught{p_4,r,c}) & \le \Singletons{u,d,4}
\end{align}
Just as for constraints (\ref{eqn:courseshed1}--\ref{eqn:courseshed3}),  
 cuts strengthening constraints (\ref{eqn:patmat1}--\ref{eqn:patmat4}) were
 described by \citet{Marecek2008TR}.
Finally, values of $\CourseRooms{}$, similar to values of $\CourseSchedule{}$,
 are forced to maxima of certain subsets of $\Taught{}$: 
\begin{align}
  \hskip 2cm \forall p \in \Periods \forall r \in \Rooms \forall c \in \Courses & 
  \hskip 2cm \Taught{p,r,c} & \le \CourseRooms{r,c} \\  
  \hskip 2cm \forall r \in \Rooms \forall c \in \Courses &
  \hskip 2cm \sum_{p \in \Periods} \Taught{p,r,c} & \ge \CourseRooms{r,c} \label{eqn:mono-end}
\end{align}
These constraints complete the formulation, which will be later referred to as 
 \textbf{Monolithic}.

\subsection{How Difficult is this for Exact Solvers?}
\label{sec:difficult}

One might hope that it is enough to pass this formulation to a modern integer 
 programming solver and wait. 
After all, integer programming has been used in timetabling ever since \citet{Lawrie1969} 
 generated feasible solutions to a school timetabling problem using a branch and bound procedure with Gomory cuts.
\citet{Tripathy1984} and \citet{Carter1989} solved instances of a course timetabling problem of up to 287 events 
 with several soft constraints using Lagrangian relaxation.
They have not, however, introduced any constraints penalising interaction between events 
 in timetables other than, of course, straightforward conflicts.
More recently, a number of modest instances of course timetabling problems have been
 tackled using off-the-shelf solvers, without introducing any new cuts  
 \citep{Dimopoulou2004,Qualizza2004,Daskalaki2004,Daskalaki2005,Mirhassani2006a}.
For instance, \citet{Daskalaki2004,Daskalaki2005} solved instances 
 of up to 211 events using ILOG CPLEX.
Of special interest are studies by \citet{AlYakoob2007} and \citet{Helber2007}, who have modelled
 a larger number of constraints. 
In the most rigorous study so far, \citet{Avella2005} presented a branch-and-cut
 solver for the Benevento Course Timetabling Problem, which forbids some interactions 
 of events in timetables other than conflicts using hard constraints. 
In this setting, \citet{Avella2005} have been able to solve instances of up to 
 233 events and 14 distinct enrolments, but conceded that application of their 
 solver to the four small instances of Udine Course Timetabling available in 
 2005 yielded ``poor results''.
In Udine Course Timetabling, such interactions are penalised by soft constraints 
 (\ref{eqn:courseshed1}--\ref{eqn:mono-end}), which (it turns out) make the problem 
 considerably more difficult.
Several integer programming formulations of the problem    
 have been studied by \citet{Marecek2007TR,Marecek2007GOR}, and more recently
 by \citet{Luebbecke2008,Lach2008}.
Both approaches can now solve the original instances of Udine Course Timetabling 
 from 2005, but optimal solutions to instances from the International Timetabling
 Competition 2007, or optima for large real-life instances of university course 
 timetabling problems are, as it seems, out of reach, so far.

 \TODO{Explained above about ``interaction between events'' -- but
 need to check got it}

In particular, it seems that it is the run-time of the linear programming (LP)
 solver that is prohibitive to progress on more difficult instances
 from the International Timetabling Competition (ITC)
 using the \textbf{Monolithic} formulation.
For relatively easier instances (comp01 and comp11 from ITC 2007), root relaxation 
 obtained from \textbf{Monolithic} takes less than thirty seconds to solve using 
 the default ILOG CPLEX 11 Dual Simplex LP Solver and the search proceeds swiftly. 
On the remaining instances, however, the root relaxation is far more expensive,
 taking up to 6489 seconds on comp07.
For complete results obtained using the CPLEX 11 Dual Simplex LP Solver, see 
 Table~\ref{tab:lp-relaxations}.  
Informal evidence suggests that run times of both the CPLEX 11 Barrier  
 and the CPLEX 11 Primal Simplex LP solvers
 are shorter, whereas run-times of either SoPlex \citep{Wunderling1996} 
 and CLP \citep{Forrest2004} are longer.
Hence, it is not particularly surprising that for most instances (all except from comp01, comp05, and comp11), CPLEX Mixed Integer Programming (MIP)
 solver using \textbf{Monolithic} and default settings does not produce any feasible solution within 40 CPU units, branching at the speed of ten to thirty nodes per hour. 
This can hardly be described as satisfactory progress.

Although the total run-time or time spent at the root node may not be of particular
 importance, as long as the instance is solved close to optimality within a reasonable 
 time limit, for example over a weekend, relying on robustness of modern solvers does not
 seem to be an option for real-life instances \citep{Rudova2007}. 
First, such instances are often many times larger than those used in the International Timetabling
 Competition 2007, including comp07. 
Second, their formulations are considerably more complex and tend to change over time, 
 whereas the behaviour of general solvers on monolithic formulations tends to be rather 
 difficult to predict,
 with run-time often rising rapidly on addition of seemingly trivial constraints.
Finally, general solvers only seldom perform in the ``anytime'' fashion of \citet{Zilberstein1996}
 on monolithic formulations,
 which would make it possible to obtain a solution whenever the solver is stopped.  
It thus makes sense to study decompositions of the problem and the performance of 
 heuristic methods based on them.

%% file: overview.tex
\section{Multiphase Exploitation of Multiple Objective-/Value-restricted Submodels (MEMOS)}
\label{sec:over}

In short, our approach can be seen as a hybridisation of loosely coupled integer
 programming solvers working on very different sub-problems of similar difficulty. 
Each solver is, time-limit permitting, exact within the given search space, and together, 
 the solvers provide heuristic solutions with global lower bounds. 
It seems, however, useful to set this description in the broader context of hybrid metaheuristics,
  even if this still seems difficult to do precisely, but concisely,
  despite the numerous attempts \citep{Celegari1999,Puchinger2005,ElAbd2005,Raidl2006}
  to develop a taxonomy of the field.

\subsection{Hybridisation}
 
Hybridisation can clearly occur in two different fashions:
\begin{itemize}
\item Algorithm hybridisation, using multiple algorithms for the same problem
\item Formulation hybridisation, using multiple subproblems or multiple levels of abstraction 
\end{itemize}
Although, in theory, hybridisation can occur in both fashions,  
 in practice, heuristics often exploit only algorithm hybridisation: 
Multiple algorithms producing solutions to the complete problem,
 with or without a feasible solution on the input. 
A typical example of algorithm hybridisation is the ``construct-improve'' scheme, 
 which is a standard practice in metaheuristic solvers.  
A constructive algorithm is used in order to produce one, or many, 
 different initial solutions, which are then improved using some variant of local search.
See the concise overview by \citet{Blum2003} for details.
Embedding of multiple primal and improvement heuristics \citep{Fischetti2003,Danna2005,Berthold2008}  
 in general integer programming solvers
 can be seen as another example of algorithm hybridisation.  
In contrast, the deliberate creation and exploitation of multiple sub-problems, 
 restricted to certain components in the objective function, with certain values fixed, 
 or both, seems underutilised.

\begin{figure}[!t]
\centering
\includegraphics[angle=270,width=0.45\textwidth]{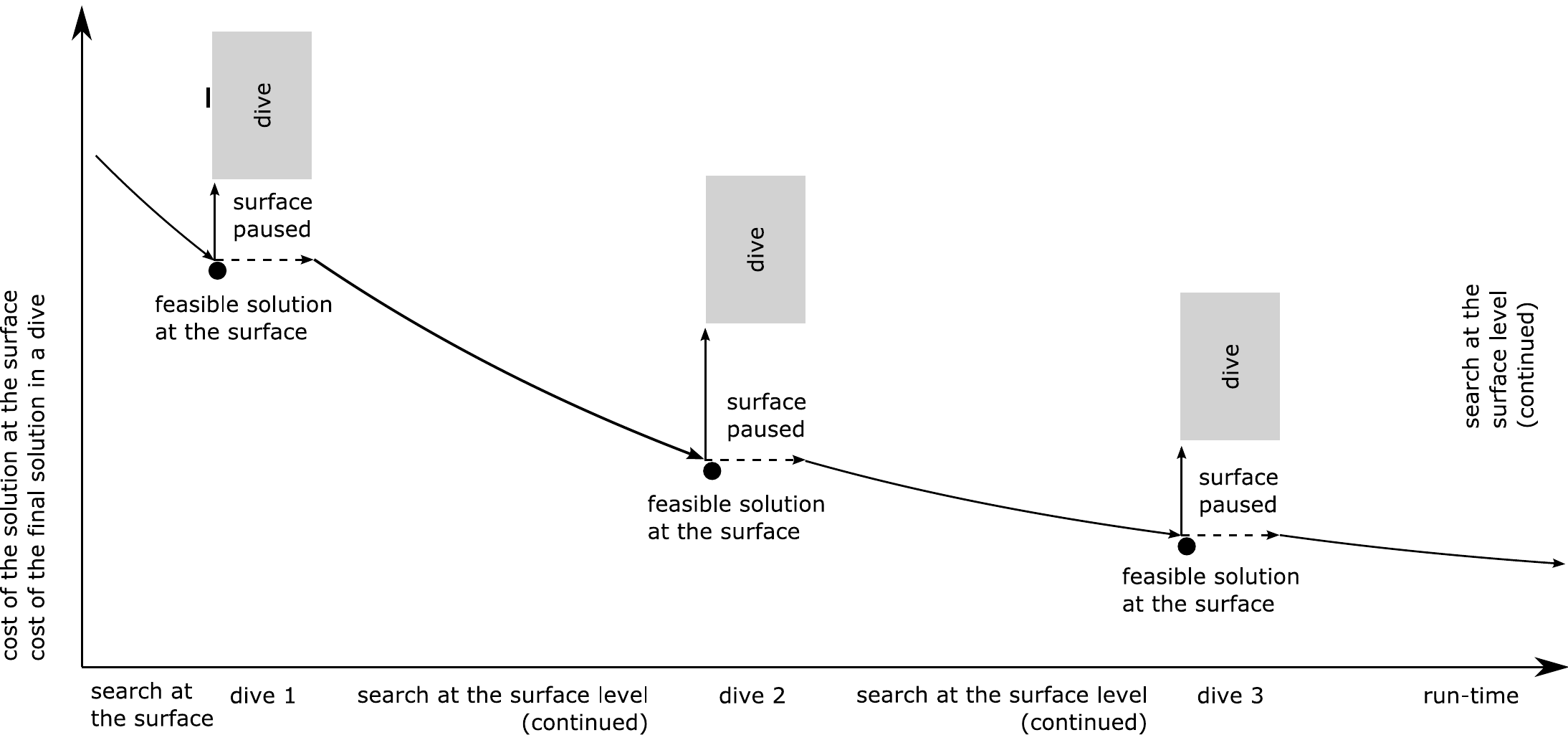} 
\caption{A schematic illustration of the data flows in the metaheuristic.}
\vskip 6mm
\label{fig:info-flow}
\end{figure}

\TODO{AJP: need to fix font sizes in all figures so that they are not significantly smaller than the text}

\subsection{Submodels}

The critical point in our approach is precisely that: sub-problems solved at
 various stages are different, although possibly overlapping.  
That is, we create sub-problems using combinations of

\begin{itemize}
 \item \textbf{Value-Restrictions}: Fix the values of a subset of the
 variables, distance from a solution with respect to Hamming distance ($L_0$ norm), 
 or sums across subsets of variables. This is a standard method to generate a neighbourhood.

 \item \textbf{Objective-Restrictions}: Suppose that we have a minimisation problem 
 and some penalty term $P^R$ is part of the objective, with associated weight $W^R$. 
 We then use one of two options:
 \begin{itemize}
 \item \textbf{Objective-Ignore-Term} We drop all consideration of the penalty -- 
 this corresponds to setting $W^R=0$. 
  \item \textbf{Objective-Fix-Term} We force $P^R=0$. This converts the penalty to a 
  hard constraint.
 \end{itemize}
 The objective restriction might loosely be thought of as the dual of
 the variable restrictions.
\end{itemize}
In any case, the intent is that the restriction will give a sub-problem that can be 
 significantly simplified compared to the full problem, and of comparable difficulty
 to other problems obtained in the decomposition.

This can be illustrated on a class of problems,
which often occurs in Scheduling and Timetabling applications, where there is
a graph colouring component expanded over another dimension. Imagine for example
the assignment of jobs to time-periods, expanded over machines, or
the assignment of events to time-periods, expanded over rooms.    
In such cases, only one part of the objective is usually related to the expanded
problem, whereas the rest is related to the graph colouring component.
This results in structures such as:
  such as:
\begin{align}
\min \left( \sum_{i = 1}^{n} c_i x_i + 
            \sum_{j = 1}^{o} d_j \max_{k \in I_j} x_k +
            \sum_{l = o+1}^{p} d_l \max(f(l, \max_{m \in I_l} x_m), 0)
     \right) \label{eqn:SummedGeneralObj} \\
     \max_{k \in I_j} x_k \le 1 \; \forall j \in \{ 1, 2, \ldots, o\} \label{eqn:SummedGeneralSumLE1} \\ 
     A x \ge b \\
     x \in \{0, 1\}^{n} \label{eqn:SummedGeneralX}
\end{align}
where $n$, $m$, $o$, and $p$ are integral constants,
 $x$ is a vector of $n$ integral decision variables, 
 $A$ is an $n \times m$ constant matrix,
 $b$, $c$, and $d$ are compatible constant vectors, and
 for each $j = \{1, 2, \ldots, p\}$, $I_j$ is a subset of $\{1, 2, \ldots, n\}$.
Notice that the second two terms in the objective function, weighted by $d$, are related only to 
 the graph colouring component, and necessitate implementation of the $\max$
 function using additional variables. 
The number of ``core'' variables $x$ can be much less than the number of additional 
 variables implementing the $\max$ function over various subsets of $x$.   
Hence, it makes sense to use the restricted objective:
\begin{align}
\min \left( \sum_{j = 1}^{o} d_j \max_{k \in I_j} x_k +
            \sum_{l = o+1}^{p} d_l \max(f(l, \max_{m \in I_l} x_m), 0)
     \right),
\end{align}
ignoring the first term in the original objective (\ref{eqn:SummedGeneralObj}).
Given the constraint matrix $A$ exhibit certain structure (\ref{eqn:SummedGeneralSumLE1}), this should 
 allow for aggregation of $n$ variables $x$ into $o$ variables $y$: 
\begin{align}
\min \left( 
     \sum_{j = 1}^{o} d_j y_j + 
     \sum_{l = o+1}^{p} d_l \max(f(l, y_{g(l)}), 0)
     \right) \label{eqn:SurfaceGeneralObj}
     A' y \ge b' \\
     y \in \{0, 1\}^{o} \label{eqn:SurfaceGeneralY}      
\end{align}
where $g$ is a suitable mapping.
Notice that some constraints may have to be relaxed in deriving the new constraints ($A' b'$) from the original ones ($A b$).
In Table~\ref{tab:intro} and below, we denote such a subproblem as ``surface''.
Such an objective-restricted subproblem, however, should solve considerably faster than
 the original problem (\ref{eqn:SummedGeneralObj}--\ref{eqn:SummedGeneralX}).

Once a solution to the objective-restricted subproblem is found, it makes sense
to resolve the original problem (\ref{eqn:SummedGeneralObj}--\ref{eqn:SummedGeneralX}) 
with the additional constraints:
\begin{align}
            \max_{k \in I_j} x_k = y_j & \;\;\; \forall j \in \{1, 2, \ldots, o\} \\
            \max(f(l, \max_{m \in J_l} x_m) = y_l & \;\;\; \forall l \in \{ o+1, o+2, \ldots, p\}.
\label{eqn:DiveGeneral} 
\end{align}
In Table~\ref{tab:intro} and below, we denote such subproblems as ``dives''. 
This should result in considerable reductions of the model in presolving,
and consequently in value-restricted subproblems which solve considerably faster than
 the original problem (\ref{eqn:SummedGeneralObj}--\ref{eqn:SummedGeneralX}).

One can also use any combination of these, and could also separately 
 treat different terms of the objective.
This is, however, limited by the availability of a decomposition into such sub-problems,
 given by the structure of a solution and the objective,
 applicable reformulations, and our ability to automate their execution.
For many ``academic'' problems there tends to be just a single term on the objective -- 
 for example, for the TSP we have only the tour length, and the objective restriction is hence 
 not relevant. 
For many real-life problems, however, it is quite likely that 
 the solution has a more complex structure, and
 there are multiple soft constraints corresponding to multiple penalty terms in the objective. 
University course timetabling \citep{Rudova2007} as opposed to pure graph colouring, 
 or vehicle routing problems \citep{Chabrier2006,Pisinger2007} as opposed
 to the travelling salesman problem, provide good examples of
 possible application areas. 
In this case ignoring or fixing some terms can give significantly easier problems, 
 yet still provide enough information to guide solution of the full problem.
As we will see, automated preprocessing in-built in modern integer programming solvers 
 is often too weak, and we do need to apply non-automatic reformulations ``by-hand''. 
Ultimately, our work is directed towards automating decompositions and reformations 
 of such problems with ``multiple soft constraints'', 
 exploiting cases in which the objectives can have differing levels of importance.

\begin{figure}[!t]
\centering
\includegraphics[angle=270,width=0.45\textwidth]{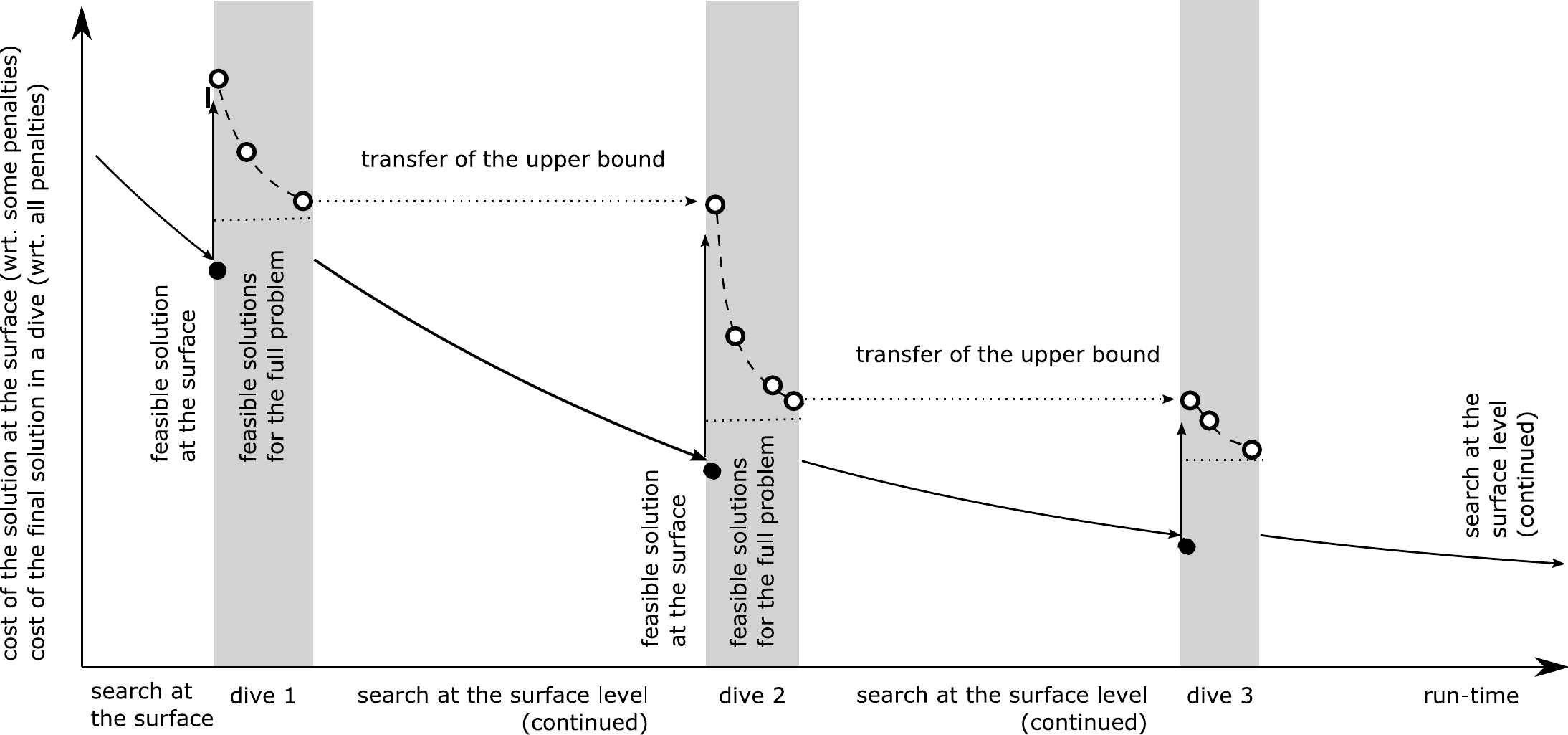} 
\caption{A schematic illustration of information flows in the anytime algorithm:
feasible solutions at the surface are transformed into neighbourhoods for dives,
executed immediately. Upper bounds (costs of best-so-far solutions) are being 
carried over from dive to dive.}
\vskip 6mm
\label{fig:info-flow-anytime}
\end{figure}

\subsection{Control Strategies}

\begin{figure}[!t]
\centering
\includegraphics[angle=270,bb=0 0 631 296,width=0.45\textwidth]{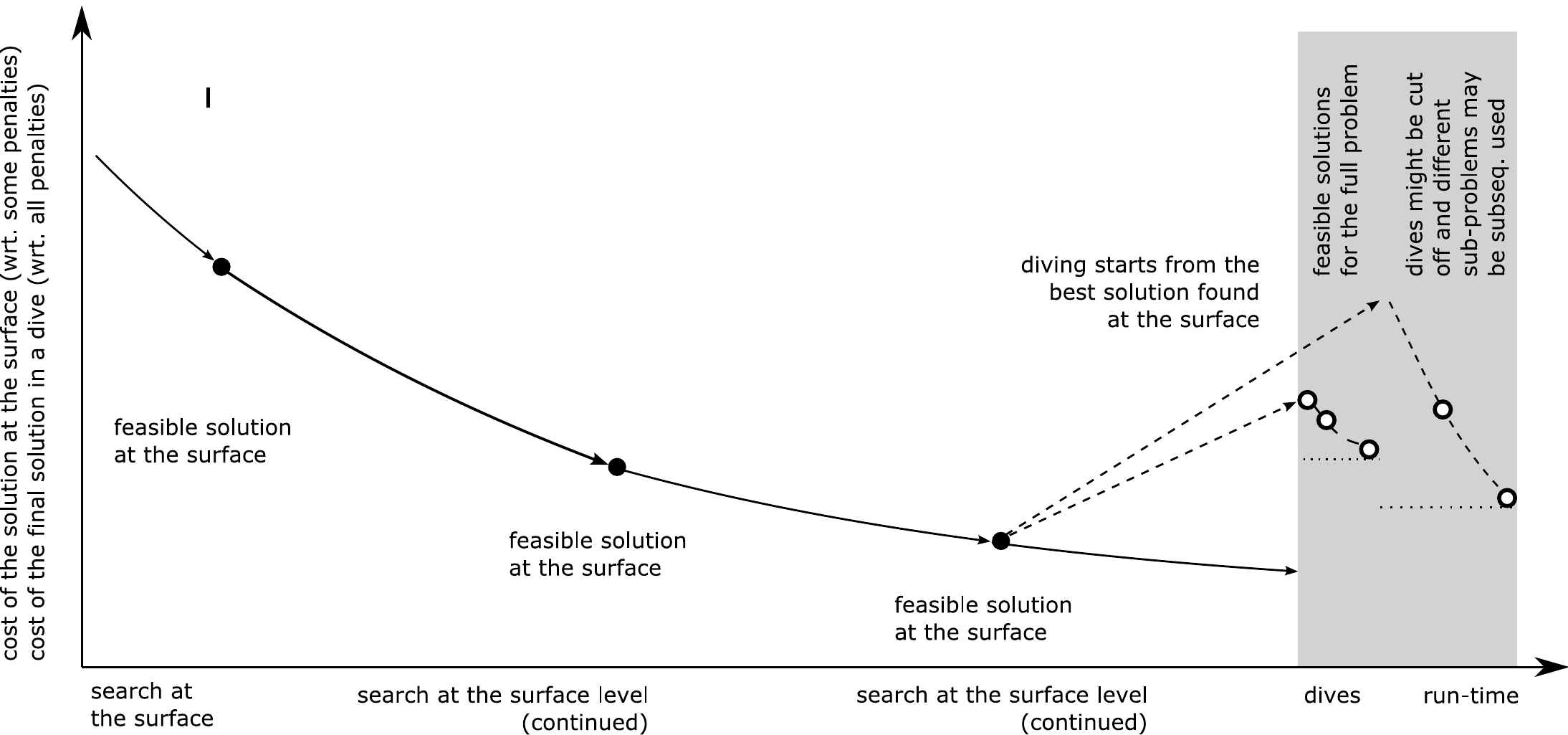} 
\caption{A schematic illustration of the contract algorithm: dives of all types are
postponed until the surface search finishes (runs out of time), when they are
carried out in the descending order of the time of discovery of the neighbourhood,
with the upper bound found from in the best neighbourhood often cutting off the
dives into other neighbourhoods as soon as their local lower bounds are established.}
\vskip 6mm
\label{fig:info-flow-contract}
\end{figure}

Another important and often neglected question in the design of hybrid metaheuristics 
 asks for the sequence of periods of searching at the surface and periods of diving
 in neighbourhoods, sometimes referred to as ``the control strategy'' \citep{Puchinger2005,Raidl2006}. 
Very often, a variant of the anytime algorithm \citep{Zilberstein1996} is used,
 perhaps because it fits well into the large neighbourhood search scheme.
There are, however, two obvious alternatives of the producer-consumer relationship:

\begin{itemize}
 \item \textbf{Anytime} algorithm: search at the surface and dives are interleaved
 and dives are perfomed as soon as suitable feasible solutions are found.
 The algorithm is parametrised with the types of dives performed, their order, and
 possibly the onset of execution of each type. 
 See also Figure~\ref{fig:info-flow-anytime}.

 \item \textbf{Contract} algorithm: search at the surface is carried out for a
 certain time and only then are the dives performed, firstly ordered by dive type
 with the least expensive dives coming first, and secondly ordered in the ascending 
 order of the cost (in minimisation problems) of the solution at the surface the 
 dive is derived from. 
 Hence, the algorithm is parametrised with the types of dives performed, the time
 limit and the proportion of the time limit spent at the surface. 
 See also Figure~\ref{fig:info-flow-contract}. 
\end{itemize}

It seems that, in practice, a time-limit is often given and considerable performance
 improvements can be gained from exploiting this in a \textbf{Contract} algorithm. 
If the neighbourhoods can be searched in the ascending order of the cost of the
 solution at the surface and the cost of the present-best solution found in dives 
 is applied as an upper bound in all dives except the first one, a considerable number 
 of dives can be cut off immediately and the algorithm needs not to be parametrised 
 with the onset of diving.
(This is suggested in Figure~\ref{fig:info-flow-contract}.)
Notice that the ascending order of the cost corresponds to the inverse order of the 
 time of discovery, if we are producing only progressively improving solutions
 at the surface.  

Generally, such Multiphase Exploitation of Multiple Objective-/value-restricted Submodels
 (MEMOS) will no longer be complete or exact. 
The intention is, however, to produce good solutions quickly. 

%% file: overview2.tex
\section{MEMOS for Udine Course Timetabling}
\label{sec:over2}

In the design of the particular MEMOS heuristic, we need to make 
 specific decisions as to which sub-problems the ``surface'' and ``dive'' components need to solve.
In doing so, it seems useful to think of the problem from the multi-objective perspective. 
In the Udine Course Timetabling problem introduced in Section~\ref{sec:problem}, 
 a given set of weights is used to linearise the penalties given by the number of
 violations of four soft constraint (``objectives'') into a single-objective problem.
The four objectives are not necessarily linked, however.
The formulation introduced in Section~\ref{sec:monolithic} models the four soft
 constraints one-by-one, 
 each objective using a separate subset of constraints and a separate subset of variables 
 derived from the main decision variables. 
Hence, it is natural to consider sub-problems arising from setting
 weights related to a subset of the four soft constraints to zero.

If some weights are set to zero, the corresponding objective can be ignored 
 and the problem formulation can be simplified. 
Ideally, one would be able to rely upon the preprocessing in-built in an 
 integer programming solver to do this.
As suggested by Table~\ref{tab:effects-objectives}, however, present-day
 general purpose integer programming solvers are not particularly effective for such major
 reformulations.
Non-automatic reformulation can, however, result in much faster relaxations, outside of
 having other desirable properties \citep{Trick2005}.
In our case, such a desirable property might be the ease of conversion of solutions 
 obtained at the ``surface'' to neighbourhoods for the dives.
As suggested in Table~\ref{tab:lp-relaxations}, a careful choice of the decomposition
 and manual reformulation can lead to the run-time of the linear programming 
 solver at the relaxation either in the surface component or within a dive
 being cut by factor of more than two thousand, compared to the ``monolithic''
 formulation.

\begin{figure}[!t]
\centering
\includegraphics[]{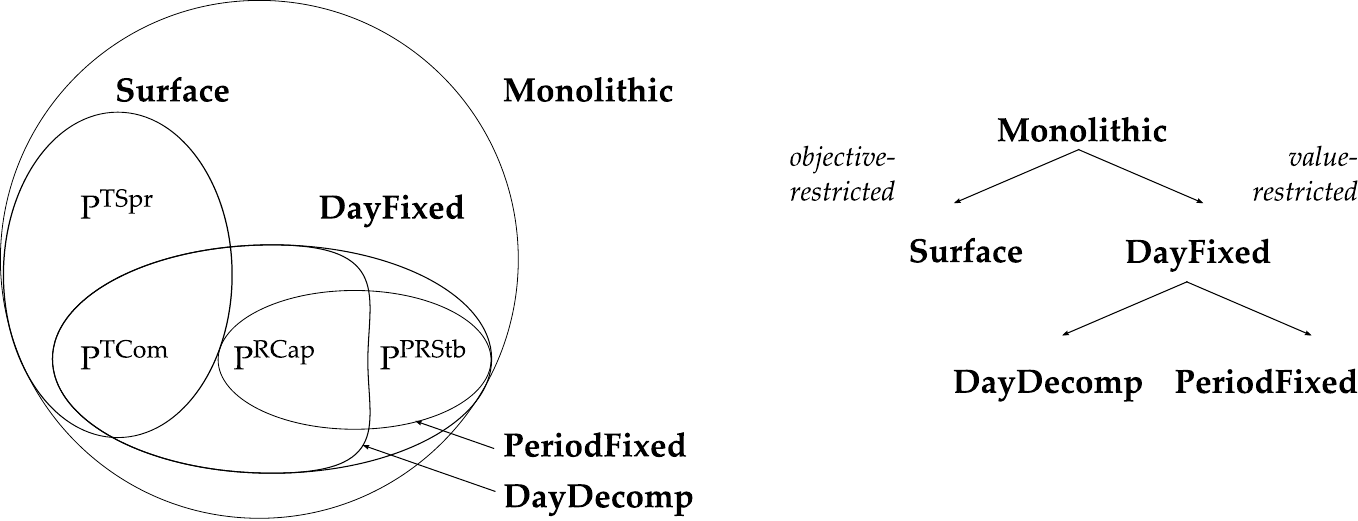} 
\caption{A schematic illustration of the penalties considered in the formulations
we use: \textbf{Monolithic} uses the complete objective function, 
\textbf{Surface} sets $\WCapacity = \WStability =  0$,
\textbf{DayFixed} has $\PSpread$ fixed,
\textbf{PeriodFixed} has $\PSpread$ and $\PCompactness$ fixed,
and \textbf{DayDecomp} has $\PSpread$ fixed and $\WStability = 0$.}
\vskip 12mm
\label{fig:hierarchy}
\end{figure}

\subsection{Objective-restricted Submodels}

One could consider generating reduced problems by setting any subset
of the weights to zero. With four penalties, there are $2^4-1=15$ 
possible non-trivial choices. However, we only use the following problem:
\begin{itemize}

 \item \textbf{Surface}: We set $\WCapacity = \WStability = 0$ so that
 room over-filling and room stability are not considered.  Since no
 explicit room assignments are required, there can be much fewer
 variables. We will see later that the ternary variables
 $\Taught{p,r,c}$ are no longer required, which gives us much shorter
 run-times of the linear programming solver. (See
 Table~\ref{tab:lp-relaxations}.) However, we add extra constraints
 that bound the number of rooms used in any single period; they impose
 that there will always be enough rooms for courses at each period,
 and so guarantee that a surface solution can always be extended by
 the dive to a feasible solution to the full problem.  However, these
 extra constraints convert the colouring to bounded colouring and make
 solving the ``surface'' search harder, both in theory and practice.

\end{itemize}

The sets of the objectives considered by each formulation are
illustrated in Figure~\ref{fig:hierarchy}.  This also serves to
emphasise the crucial point that the methods we use are intended for
the case of multiple terms in the objective, or equivalently, for
multiple classes of relatively-complex soft constraints. 

One can reasonably ask why we use this particular split between
surface and dive models.
Firstly, the graph colouring component is important in the instances 
 under consideration. In Table~\ref{tab:instances}, we can see
that the ratio of events to available room slots is typically 60-90\%
which is quite high, so it is reasonable that the bounded colouring
problem is hard so solve, and so ought to be solved first. That is,
the bounded colouring needs to be done at the surface level.  Also,
Table~\ref{tab:effects-objectives} gives optimal objective vectors for
different weight vectors. (It does this for the small instance comp01
that happens to be fairly easy to solve exactly, and uses the
monolithic IP formulation detailed in the previous section.) We see that
the time-related penalties ($\PSpread$ and $\PCompactness$) make the 
largest contributions to the final objective, and so it is natural 
to address these first.
Limited experimentation has been conducted with solvers where $\WCapacity = \WStability = \WCompactness = 0$.
As is suggested by Table~\ref{tab:effects-objectives}, the solver at
 the surface is not considerably faster, when only automatic reductions
 are considered.
Rather surprisingly, the solver is not considerably faster, however,
 even when non-automatic reformulations are applied, and these gains
 do not offset the need to spend more time diving. (See Section~\ref{sec:yet-better}.)
There are also alternatives exploiting interchangeable rooms.

 \TODO{ It would be nice to know the (bounded) chromatic numbers of
 the event graphs. Any chance the reversible clique formulation of
 purely the graph colouring can answer this? Probably not? except for
 comp01?}

\begin{table}[!t]
\centering
\renewcommand{\arraystretch}{1.3}
\caption{The effects of disabling various subsets of soft constraints by setting 
the corresponding weights in the monolithic model to zero on CPLEX 10 run-time for 
instance comp01. Only the default in-built pre-processing and cuts are used
and only the best feasible solution obtained within one hour per run is presented.
Notice that when a weight is set to zero, the associated penalty is typically
high.}
\label{tab:effects-objectives}
\vskip 3mm
\input{tables/effects-objectives}
\vskip 3mm
\caption{The effects of varying the weight $\WSpread$ on CPLEX 10 run-time for 
the monolithic model and instance comp01. }
\vskip 3mm
\input{tables/effects-objectives2}
\vskip 3mm
\end{table}

\subsection{Exploiting Interchangeable Rooms: ``Multi--rooms''}


Suppose that we have two rooms of exactly the same size. Then the only
way that the rooms are not interchangeable is because of the room stability
constraint.
If we take a solution and swap the rooms at any single time period
then no constraints are broken and no objectives affected, except
potentially the room stability.  Hence, if we are using a formulation,
such as the surface, that does not care about room stability, then we
can treat the rooms as interchangeable. Thus, we could replace the 2
standard rooms with a single room with multiplicity, which we call a
``multi-room''. 

A multi-room of size ``($m$,$a$)'' will have multiplicity $m$, 
capacity $a$, and will function exactly as $m$ separate indistinguishable
rooms. That is: It can accommodate $m$ separate events of sizes up to $a$ each,
without any overflow.
This mild extension is straightforward, but has the advantage that we
then have the option to replace the $r$ index of the $\Taught{p,r,c}$
variables by an index into a set of multi-rooms.
Any solution using multi-rooms can be translated into constraints on
original rooms that define a neighbourhood in the full problem. 
Within this neighbourhood, there will always be a feasible
solution, and we can easily find the one minimising $\PStability$,
 the penalty for ``room stability''.

Here we use multi-rooms only as a device to reduce the number of
variables and remove symmetries that would enlarge the search
space. However, it is also reasonable that multi-rooms could be
directly useful. Specifically, in practice, the room stability might
well not be quite so strict as the Udine timetabling instances
demand. In some teaching facilities, there could well be a suite of
teaching rooms that are all of the same size, with the same facilities,
and which are located very close to each other. In such a case, it could well be
that there really would be no need for a room stability penalty for
mixed use. In this case, they could well be treated exactly as a
single multi-room with appropriate multiplicity.


Generally, only a few rooms do have exactly the same sizes. However,
it is often the case that there are many rooms of roughly the same size. 
To exploit this situation, note that if we join two rooms of different
sizes into a single multi-room then it can affect the room capacity
objective, but again it will {\em not} affect the hard constraints.
Hence, we do allow the creation of relaxations in which various sets of
rooms are joined into associated multi-rooms with capacity the size of
the largest original room.
Specifically, if we select a set of multi-rooms with capacities $\{c_i |
i = 1 \ldots \abs{\Rooms}\}$ then we can replace them with a multi-room of size $(m,
\max c_i \}$.
An intermedaite case would be to divide the rooms into two
multi-rooms, according to whether they are smaller or not than some
selected intermediate size. We will introduce such a neighbourhood generator,
denoted by \textbf{Surface2}, in Section~\ref{sec:subproblems}.

\subsection{Value-restricted Submodels}

For the dives, we use value-restricted submodels of the \textbf{Monolithic} model.
Two pure value-restricted submodels are defined by restricting the times of 
courses based on the solution obtained at the surface level:

\begin{itemize}

 \item \textbf{PeriodFixed} dives: The periods of all the courses are fixed
 to be the same as obtained from the surface solution.

 \item \textbf{DayFixed} dives: The days of all the courses are fixed, but
 not the explicit period within the day. That is, for each course
 the assigned ``period'' is relaxed to just give a day assignment.

\end{itemize}

In both pure value-restricted subproblems, the spread of events of a course over distinct week-days 
 is entirely fixed, and $\WSpread$ hence becomes irrelevant.
\textbf{PeriodFixed} dives also fix time compactness penalty $\PCompactness$, and thus only 
 room assignment remains to be found, minimising the room related penalties ($\PCapacity$ and $\PStability$).  
In \textbf{DayFixed} dives, the compactness penalty $\PCompactness$ can still potentially be improved.
A middle ground between these two approaches is offered by value- and objective-restricted
submodels. 

\subsection{Value- and Objective-restricted Submodels}

Two obvious value- and objective-restricted submodels can be defined by restricting the 
times of courses based on the solution obtained at the surface level, and not 
considering $\PStability$ or by forcing $\PStability$ to zero:

\begin{itemize}

 \item \textbf{DayDecomp} dives: $\PStability$ is not considered and the days of all 
 the courses are fixed, but not the explicit period within the day. That is,  
 for each course, the assigned ``period'' is relaxed to just give a day assignment
 and the lowest possible penalty for time compactness and room capacity is sought.

 \item \textbf{DayFixedZero$\PStability$} dives: $\PStability$ is forced to be zero and the 
 days of all the courses are fixed, but not the explicit period within the day. That 
 is, for each course, the assigned ``period'' is relaxed to just give a day assignment
 and an assignment with zero penalty for room stability is sought.
\end{itemize}

These two value- and objective-restricted submodels have the advantage that they
 reduce the links between variables representing individual days, which translates
 into matrices closer to the block structure and improvements in the performance 
 of the diving integer programming solver.   
Many other subproblems, most notably variants of \textbf{PeriodFixed} and \textbf{DayFixed} 
 dives using stochastic ruining \citep{Schrimpf2002,IBM2002}, are of course possible, 
 but are not studied here.
Some experience with implementing and testing the presented alternatives empirically is 
 described in Section~\ref{sec:results}.  

\subsection{Control Strategies}

Finally, we have designed both \textbf{Anytime} and \textbf{Contract} control strategies
 for Udine Course Timetabling, 
 using \textbf{PeriodFixed} dives followed by \textbf{DayFixed} dives.
The details of their functioning is outlined in Figure~\ref{fig:pseudocode}.
\textbf{Contract} algorithms turned out to perform better, as expected, especially
  when multiple solutions can be found at \textbf{Surface} within the given time limit. 
The only difficulty seems to be the choice of the partition of short time limits 
  between \textbf{Surface} and dives in \textbf{Contract} algorithms.
When a time limit is given, and is not too strict, however, there seem to be no reasons to
  use \textbf{Anytime} strategies.

\begin{figure}[!t]
\centering
\begin{Verbatim}
class AnytimeStrategy implementing Strategy:
  constructor(config):
    keep track of config
    initialise bestSoFar to an error value
  onFeasibleSolution():
    foreach type in config.getDiveTypes():
      if time > config.getAnytimeDivingOnset(type):
        Neighbourhood def = getNeighbourhood(type)
        update bestSoFar with 
          dive(def, config.getDiveTimeLimit(type))
  onTimeout():
    return bestSoFar

class ContractStrategy implementing Strategy:
  constructor(config):
    keep track of config
    initialise a stack of neighbourhoods n[t] 
      for each t in config.getDiveTypes()
  onFeasibleSolution():
    foreach type in config.getDiveTypes():
      fixDayNeighbs.push(getNeighbourhood(type))
  onTimeout():
    initialise bestSoFar to an error value
    foreach type in config.getDiveTypes():
      c = 0
      while time > config.getTimelimit() && 
          ! n[type].empty() && c < config.getRunLimit(type):
        update bestSoFar with 
          dive(n[type].pop(), config.getDiveTimeLimit(type))
        c += 1

class SampleAnytimeSolver():
  solve():
    config.setDiveTypes([FixPeriod, FixDay])
    foreach type in config.getDiveTypes():
      come up with a suitable time limit for type
      config.setDiveTimeLimit(type, limit)
    load an instance
    initialise a solver for Surface, instance, and config
    install AnytimeStrategy(config) as a solver callback
    try solving Surface

class SampleConstractSolver():
  solve(timeLimit):
    config.setTimeLimit(timeLimit)
    config.setDiveTypes([FixPeriod, FixDay])
    config.setDiveTimeLimit(FixPeriod, timeLimit/50)
    config.setDiveTimeLimit(FixDay, timeLimit/5)
    foreach type in config.getDiveTypes():
      config.setRunLimit(type, 3)
    load an instance
    initialise a solver for Surface, instance, and config
    install ContractStrategy(config) as a solver callback
    try solving Surface for timeLimit/3
\end{Verbatim}
\caption{Pseudocode of the control strategies in use}
\vskip 6mm
\label{fig:pseudocode}
\end{figure}

%% file: tables/effects-objectives.tex
\begin{tabular}{cccc|ccrr|cccc|c}
\rotatebox{90}{$\WCapacity$} & \rotatebox{90}{$\WSpread$} & \rotatebox{90}{$\WCompactness$} & \rotatebox{90}{$\WStability$} & LP Matrix & \rotatebox{90}{Barrier LP} \rotatebox{90}{Runtime} & \rotatebox{90}{Node with} \rotatebox{90}{last feas. sol.} & \rotatebox{90}{Default IP} \rotatebox{90}{Runtime} & \rotatebox{90}{$\PCapacity$} & \rotatebox{90}{$\PSpread$} & \rotatebox{90}{$\PCompactness$} & \rotatebox{90}{$\PStability$} & Obj.\\ \hline
0 & 0 & 0 & 0 & $ 6639 \times 5407 $ & 44.08 s &0 & 66~s & 2294 & 30 & 350 & 93 & 0 \\ 
\hline
0 & 0 & 0 & 1 & $ 6639 \times 5407 $ & 43.66 s &370 & 1066~s & 2706 & 30 & 350 & 0 & 0 \\ 
0 & 0 & 2 & 0 & $ 7059 \times 5757 $ & 61.24 s &140 & 884~s & 2107 & 30 & 0 & 77 & 0 \\ 
0 & 5 & 0 & 0 & $ 6665 \times 5435 $ & 40.44 s &0 & 148~s & 2387 & 0 & 350 & 92 & 0 \\ 
1 & 0 & 0 & 0 & $ 6639 \times 5407 $ & 21.68 s &0 & 62~s & 4 & 30 & 350 & 44 & 4 \\ 
\hline
0 & 0 & 2 & 1 & $ 7059 \times 5757 $ & 67.73 s &617 & 1033~s & 2874 & 30 & 0 & 0 & 0 \\ 
0 & 5 & 0 & 1 & $ 6665 \times 5435 $ & 39.41 s &402 & 1554~s & 2023 & 0 & 350 & 0 & 0 \\ 
0 & 5 & 2 & 0 & $ 7085 \times 5785 $ & 66.18 s &90 & 881~s & 2365 & 0 & 0 & 78 & 0 \\ 
1 & 0 & 0 & 1 & $ 6639 \times 5407 $ &  9.97 s &180 & 521~s & 4 & 30 & 350 & 1 & 5 \\ 
1 & 0 & 2 & 0 & $ 7059 \times 5757 $ & 21.21 s &124 & 518~s & 4 & 30 & 0 & 31 & 4 \\ 
1 & 5 & 0 & 0 & $ 6665 \times 5435 $ & 10.01 s &0 & 52~s & 4 & 0 & 350 & 44 & 4 \\ 
\hline
0 & 5 & 2 & 1 & $ 7085 \times 5785 $ & 39.50 s &1964 & 1050~s & 1672 & 0 & 0 & 0 & 0 \\ 
1 & 0 & 2 & 1 & $ 7059 \times 5757 $ & 21.40 s &487 & 1416~s & 4 & 30 & 0 & 1 & 5 \\ 
1 & 5 & 0 & 1 & $ 6665 \times 5435 $ & 7.96 s  &269 & 651~s & 4 & 0 & 350 & 1 & 5 \\ 
1 & 5 & 2 & 0 & $ 7085 \times 5785 $ & 22.35 s &145 & 395~s & 4 & 0 & 0 & 43 & 4 \\ 
\end{tabular}

%% file: tables/effects-objectives2.tex
\begin{tabular}{cccc|ccrr|cccc|c}
\rotatebox{90}{$\WCapacity$} & \rotatebox{90}{$\WSpread$} & \rotatebox{90}{$\WCompactness$} & \rotatebox{90}{$\WStability$} & LP Matrix & \rotatebox{90}{Barrier LP} \rotatebox{90}{Runtime} & \rotatebox{90}{Node with} \rotatebox{90}{last feas. sol.} & \rotatebox{90}{Default IP} \rotatebox{90}{Runtime} & \rotatebox{90}{$\PCapacity$} & \rotatebox{90}{$\PSpread$} & \rotatebox{90}{$\PCompactness$} & \rotatebox{90}{$\PStability$} & Obj.\\ \hline
1 & 2 & 2 & 1  & $ 7085 \times 5785 $ & 19.19 s &1053 & 1205~s & 4 & 0 & 0 & 1 & 1 \\ 
1 & 4 & 2 & 1  & $ 7085 \times 5785 $ & 19.19 s &405 & $> 3600$~s & 4 & 0 & 0 & 2 & (6) \\ 
1 & 8 & 2 & 1  & $ 7085 \times 5785 $ & 19.49 s &1420 & $> 3600$~s & 5 & 0 & 0 & 1 & (6) \\ 
1 & 16 & 2 & 1 & $ 7085 \times 5785 $ & 19.23 s &4822 & $> 3600$~s & 4 & 0 & 0 & 2 & (6) \\ 
1 & 32 & 2 & 1 & $ 7085 \times 5785 $ & 19.06 s &586 & $> 3600$~s & 4 & 0 & 0 & 2 & (6) \\ 
1 & 64 & 2 & 1 & $ 7085 \times 5785 $ & 19.39 s &3803 & 2387~s & 4 & 0 & 0 & 1 & 5
\end{tabular}

%% file: ip-subproblems.tex
\section{Non-Automatic Re-Formulations of the Subproblems}
\label{sec:subproblems}
As has been described in Section~\ref{sec:over}, at the ``surface'' level of the
 heuristic for Udine Course Timetabling, 
 there is an objective-restricted neighbourhood generator trying to find an assignment 
 of events to periods, such that at most a given number of events 
 takes place at any given period, but disregarding other issues of assignment 
 of rooms, such as room capacities.  
Feasible solutions of the assignment of events to periods are relaxed into value-restricted neighbourhoods
 given by the assignment of events to days or periods.
It would obviously be possible to use the monolithic formulation
 with weights $(\WCapacity \WSpread \WCompactness \WStability) = (0, 5, 2, 0)$
 at the ``surface'' level, and to rely on the automatic pre-solving routines in a modern integer 
 programming solver.
As will be shown in Section~\ref{sec:results}, however, improvements in the run-time of the
 linear programming solver, which determine the run-time of the integer programming solver,
 of several orders of magnitude can be achieved by non-automatic reformulation.

\subsection{The Search for Good Neighbourhoods}
In the formulation that we denote as \textbf{Surface}, 
 the ``core'' decision variables $\SetTimes{}$ 
 are accompanied by dependent variables $\CourseSchedule{}$, $\CourseMinDaysViolations{}$, 
 and $\Singletons{}$, introduced previously:
\begin{itemize} 
\item $\SetTimes{}$ are binary decision variables indexed with periods and courses. 
Their values are constrained so that in any feasible solution, course $c$ should be 
 taught at period $p$, if and only if $\SetTimes{p,c}$ is set to one.
\end{itemize}

The objective function can be expressed as:
\begin{align}  
  \min \; \WCompactness \sum_{u \in \Curricula} \sum_{d \in \Days} \sum_{s \in \SingletonChecks} \Singletons{u,d,s} + \; \WSpread \sum_{c \in \Courses} & \CourseMinDaysViolations{c} \notag
\end{align}

Hard constraints can be formulated similarly to those in \textbf{Monolithic}:
\begin{align}
  \label{eqn:ng:hardFirst}
  \hskip 2cm \forall c \in \Courses \hskip 2cm & \sum_{p \in \Periods} & \SetTimes{p,c} & = \HasEventsCount{c} \\
  \label{eqn:ng:hardThird}
  \hskip 2cm \forall p \in \Periods \forall u \in \Curricula \hskip 2cm & \sum_{c \in \CurriculumHasCourses{u}} & \SetTimes{p,c} & \le 1 \\
  \label{eqn:ng:hardFourth}
  \hskip 2cm \forall p \in \Periods \forall t \in \Teachers \hskip 2cm & \sum_{c \in \TeacherHasCourses{t}} & \SetTimes{p,c} & \le 1 \\
  \label{eqn:ng:hardSecond}
  \hskip 2cm \forall p \in \Periods \hskip 2cm & \sum_{c \in \Courses} & \SetTimes{p,c} & \le \abs\Rooms \\
  \label{eqn:ng:hardFifth}
  \hskip 2cm \forall \left\langle {c, p} \right\rangle \in \Deprecated \hskip 2cm & & \SetTimes{p,c} & = 0  
\end{align}
Notice that constraint (\ref{eqn:ng:hardSecond}) is the only mention of rooms
 in this formulation of neighbourhood definition.
It makes the number of rooms used in any period smaller than $\abs\Rooms$, 
 which corresponds to making the colouring bounded. 
It would be possible, for example, to constrain the number of large courses taught 
 in any one period to be less than the number of large rooms, 
 with suitable definitions of ``large''.
Similar constraints, however, do not seem to improve the quality of neighbourhoods
 significantly.

By minor changes of constraints (\ref{eqn:courseshed1}--\ref{eqn:courseshed3}) in the monolithic formulation, 
 the array $\CourseMinDaysViolations{}$ can be constrained using:
\begin{align}
  \label{eqn:ng:cttFirst}
  \hskip 2cm \forall c \in \Courses \forall d \in \Days \forall p \in \HasPeriods{d} \hskip 2cm &  
  \SetTimes{p,c} & \le \CourseSchedule{c,d} \\  
  \label{eqn:ng:cttSecond}
  \hskip 2cm \forall c \in \Courses \forall d \in \Days \hskip 2cm & 
  \sum_{p \in \HasPeriods{d}} \SetTimes{p,c} & \ge \CourseSchedule{c,d} \\
  \label{eqn:ng:cttThird}
  \hskip 2cm \forall c \in \Courses \hskip 2cm &
  \sum_{d \in \Days} \CourseSchedule{c,d} & \ge \HasMinDays{c} - \CourseMinDaysViolations{c}
\end{align}
Inequalities \ref{eqn:ng:cttFirst} and \ref{eqn:ng:cttSecond} in effect aggregate
 values from $\SetTimes{}$ into $\CourseSchedule{}$.
Inequality~\ref{eqn:ng:cttThird} then calculates the number of days the instruction
 is short of the recommended value in $\HasMinDays{}$.

Patterns can be penalised again using the natural formulation, used already in
 constraints (\ref{eqn:patmat1}--\ref{eqn:patmat4}) in the monolithic model.
 For instances with four periods per day, we obtain: 
\begin{align}
  \label{eqn:ng:patmat1}
  \forall u \in \Curricula d \in \Days \forall \left\langle {p_1, p_2, p_3, p_4} \right\rangle \in \HasPeriods{d} &    
  \sum_{c \in \CurriculumHasCourses{u}}  & 
  (\SetTimes{p_1,c} - \SetTimes{p_2,c})  & \le \Singletons{u,d,1}
\\
  \label{eqn:ng:patmat2}
  \forall u \in \Curricula d \in \Days \forall \left\langle {p_1, p_2, p_3, p_4} \right\rangle \in \HasPeriods{d} &    
  \sum_{c \in \CurriculumHasCourses{u}}  & 
  (\SetTimes{p_4,c} - \SetTimes{p_3,c})  & \le \Singletons{u,d,2}
\\
  \label{eqn:ng:patmat3}
  \forall u \in \Curricula d \in \Days \forall \left\langle {p_1, p_2, p_3, p_4} \right\rangle \in \HasPeriods{d} &    
  \sum_{c \in \CurriculumHasCourses{u}} & 
  (\SetTimes{p_2,c} - \SetTimes{p_1,c} - \SetTimes{p_3,c})  & \le \Singletons{u,d,3}
\\
  \label{eqn:ng:patmat4}
  \forall u \in \Curricula d \in \Days \forall \left\langle {p_1, p_2, p_3, p_4} \right\rangle \in \HasPeriods{d} &    
  \sum_{c \in \CurriculumHasCourses{u}}  & 
  (\SetTimes{p_3,c} - \SetTimes{p_2,c} - \SetTimes{p_4,c})  & \le \Singletons{u,d,4}
\end{align}
This formulation of \textbf{Surface} can obviously also be strengthened by addition of cuts, as 
  described by \cite{Marecek2008TR}.

\subsection{Better Neighbourhoods?}
\label{sec:yet-better}

The approach outlined above is not the only possible neighbourhood generator, although 
 the range of possible formulations is somewhat limited by the choice of decision variables. 
Considering further soft constraints would necessitate the addition of a considerable
 number of variables, resulting in very slow LP relaxations.
(See Table \ref{tab:lp-relaxations}
 for an illustration of the effect and Section~\ref{sec:results} for discussion.)    
Replacing the array $\SetTimes{}$ with an array indexed with 
 courses and days would make it impossible to implement the graph colouring component.
The only option left, hence, seems to be the consideration of only a single soft constraints,
 setting $(\WCapacity \WSpread \WCompactness \WStability) = (0, 5, 0, 0)$. 
This enables the removal of $\Singletons{}$ and constraints (\ref{eqn:ng:patmat1}--\ref{eqn:ng:patmat4}).
It seems, however, that using weights $\Weights = (0, 5, 0, 0)$ makes it necessary to search
 larger neighbourhoods in dives (see below for \textbf{DayFixed}),
 and results in worse performance on both short and long runs.
Alternatively, we can exploit interchangeable rooms.

As has been suggested in Section~\ref{sec:over2},
\textbf{Surface} can be thought of as the extreme case of room aggregation, in which all
the $\abs{\Rooms}$ rooms are joined into a single multi-room of multiplicity $\abs{R}$
and capacity the size of the largest room. 
In all these instances, the
resulting capacity is always larger than the largest event, and so the
multi-room multiplicity just gives the bound on the number of events
per time-slot.
An intermedaite case would be to divide the rooms into two
multi-rooms, according to whether they are smaller or not than some
selected intermediate size. 
In a neighbourhood generator denoted \textbf{Surface2}, the threshold room size is just taken to be the median
of all the room sizes. This keeps the number of variables down, but
still adds some restrictions so that larger events are likely to be
evenly spread out over the periods.

More formally, given the set $\MRooms$ of pairs $(m, a)$, representing multiplicity and capacity 
 of a multi-room, the formulation we denote \textbf{Surface2} uses 
 the ``core'' decision variables $\Taught{}'$,
 which differ from $\Taught{}$ only in that they are indexed with $\MRooms$ instead of $\Rooms$,
 together with dependent variables $\CourseSchedule{}$, $\CourseMinDaysViolations{}$, 
 and $\Singletons{}$, introduced previously, 
 and together with $\CourseRooms{}'$, which are similar to
 $\CourseRooms{}$, only indexed with $\MRooms$ instead of $\Rooms$.
The objective function can be expressed as:
\begin{align}  
  \min & \; \WCapacity \sum_{(m,a) \in \MRooms} \sum_{p \in \Periods} \sum_{\substack{c \in \Courses \\ \HasStudents{c} > a}} 
       \Taught{p,(m,a),c}' \; (\HasStudents{c} - a)  
     + \; \WCompactness \sum_{u \in \Curricula} \sum_{d \in \Days} \sum_{s \in \SingletonChecks} \Singletons{u,d,s} \notag \\ 
     + & \; \WSpread \sum_{c \in \Courses} \CourseMinDaysViolations{c} 
     + \; \WStability \sum_{c \in \Courses} \left( {\left( {\sum_{r \in \MRooms} \CourseRooms{r,c}'} \right) - 1} \right) \notag
\end{align}
Hard constraints can also be formulated similarly to those in \textbf{Monolithic} (\ref{eqn:hardFirst}--\ref{eqn:hardFifth}):
\begin{align}
    \hskip 2cm \forall c \in \Courses \hskip 2cm & \sum_{p \in \Periods} \sum_{r \in \MRooms} & \Taught{p,r,c} & = \HasEventsCount{c} \\
    \hskip 2cm \forall p \in \Periods \forall (m,a) \in \MRooms \hskip 2cm & \sum_{c \in \Courses} & \Taught{p,r,c} & \le m \\  
    \hskip 2cm \forall p \in \Periods \forall c \in \Courses \hskip 2cm & \sum_{r \in \MRooms} & \Taught{p,r,c} & \le 1 \\
    \hskip 2cm \forall p \in \Periods \forall t \in \Teachers \hskip 2cm & \sum_{r \in \MRooms} \sum_{c \in \TeacherHasCourses{t}} & \Taught{p,r,c} & \le 1 \\ 
    \hskip 2cm \forall p \in \Periods \forall u \in \Curricula \hskip 2cm & \sum_{r \in \MRooms} \sum_{c \in \CurriculumHasCourses{u}} & \Taught{p,r,c} & \le 1 \\
    \hskip 2cm \forall \left\langle {c, p} \right\rangle \in \Deprecated \hskip 2cm & \sum_{r \in \MRooms} & \Taught{p,r,c} & = 0  
\end{align}
The remaining constraints of \textbf{Monolithic} can be used in \textbf{Surface2}, when
 $\Taught{}'$ replaces $\Taught{}$, $\CourseRooms{}'$ replaces $\CourseRooms{}$,
 and $\MRooms$ replaces $\Rooms$.

\subsection{The Diving} 

When an integer feasible solution is found at \textbf{Surface}, it 
 waits to be translated to a solution of the Udine Course Timetabling Problem.
There are at least two obvious restrictions of the monolithic formulation, 
 which we refer to as \textbf{PeriodFixed} and \textbf{DayFixed} dives
 in Section~\ref{sec:over}.
It is clearly possible to constraint the monolithic formulation into a \textbf{PeriodFixed} dive 
 by the addition of constraints: 
\begin{align}
  \label{eqn:transfer2}
  \hskip 2cm \forall c \in \Courses \forall p \in \Periods \hskip 2cm &  
  \sum_{r \in \Rooms}
  \Taught{p,r,c} = \SetTimes{p,c}  
\end{align}
On the instances from the International Timetabling Competition 2007, it is often possible
 to find optima for \textbf{PeriodFixed} dives within seconds. As will be described in
 Section~\ref{sec:results}, however, the quality of solutions obtained this way varies
 widely from neighbourhood to neighbourhood.  

Alternatively, it is possible to dive into neighbourhoods defined by assignment
 of courses to days. In order to do so, values of the array $\SetDays{}$ indexed
 with courses and days are pre-computed outside of the solver as:
\begin{align}
  \hskip 2cm \forall p \in \Periods \forall d \in \Days \hskip 2cm & 
   \SetDays{d,c} = \sum_{p \in \HasPeriods{d}} \SetTimes{p,c} \notag  
\end{align}
Then, it is possible to constraint the monolithic formulation using 
\begin{align}
  \label{eqn:transfer}
  \hskip 2cm \forall c \in \Courses \forall d \in \Days \forall p \in \HasPeriods{d} \hskip 2cm &
  \sum_{r \in \Rooms} \Taught{p,r,c} = \SetDays{d,c}  
\end{align}
Presolving routines of modern general purpose integer programming solvers are typically
 able to shrink integer programming instances obtained in \textbf{DayFixed} dives by the factor 
 of ten or more. 
Subsequently, optima within such neighbourhoods are found within minutes for the instances
 from the International Timetabling Competition. 
For details, see Section~\ref{sec:results}.

\subsection{Why Not Better Diving?}

A middle ground between these two solutions is represented by ``ruining'' \citep{Schrimpf2002}
 into the assignment of events to days only those elements $\SetTimes{}$, which seem to contribute
 to the objective function, or to its factor $\PCompactness$.
The first option is easy to implement, as modern integer programming solvers expose
 the pseudo-cost of individual variables, but it does not seem to be particularly useful
 for instances from the International Timetabling Competition 2007.
The implementation of the second option involves a certain computational effort, but it might
 prove useful in some instances. \TODO{IMPLEMENT BEFORE YOU SUBMIT!} 
Both kinds of ``alternative dives'' are disregarded in the remainder of the paper, and might be
 the focus of future work.
\NOTE{JXM: it is unclear
 whether they are not covered by an IBM patent \citep{IBM2002}.}

%% file: related.tex
\clearpage
\section{Related Research Issues: A Brief Discussion}
\label{sec:related}

It should be noted that the components of this approach draw upon a rich history
 of work in this area.
%
Decompositions have been used in integer programming for
 a long time \citep[provide a survey]{Ralphs2006MP} and the effects of
 non-automatic reformulations are also well known \citep[for example]{Williams1978,Nemhauser1993}. 
In the timetabling community, the ``times first, rooms second'' decomposition
 is a standard procedure. See \citet{Burke1999a} for references. 
Decompositions using integer programming are less common, although
  \cite{Lawrie1969}, for instance, used a form of column generation as early as 1969.
Decompositions providing lower bounds are yet less common,
  and often limited in constraints they allow.
\cite{Carter1989} used only variables indexed with rooms and courses, which prohibits
 formulation of period-related soft constraints, but imposes a variant of room
 stability via cuts.   
More recently, \citet{Daskalaki2005} used ``times first, rooms second''
 with integer programming models solved at both stages,
 although, in our terms, they use only a single type of dives, anytime control 
 strategy, and evaluate the performance only on two small instances.
Yet more recently, \citet{Luebbecke2008} have also developed a similar approach, 
 independently of the present authors.\footnote{
Results of \citet{Luebbecke2008} were first submitted in February 2008 to 
 Practice and Theory of Automated Timetabling, a conference held in Montr{\' e}al, 
 Canada, in August 2008. 
The approach of the present authors was first presented at the Mixed Integer Programming
 Workshop in August 2007.
}
They use a rather different and interesting surface component, 
 based on their studies of the Partial Transversal Polytope \citep{Lach2008}.
Their very respectable results are discussed below, in Section~\ref{sec:results},
 and are included in Figure~\ref{fig:results} for comparison.
In machine scheduling, such approaches are often termed ``machine aggregation 
  heuristics'' \cite{Leung2004} and it is not too difficult to translate rooms
  to machines and vice versa.
We have, however, been unable to find a more general two-staged decomposition 
 methodology using integer programming at both stages, comparable to the one presented here.
Overall, we heartily recommend both papers on decompositions of timetabling 
 \citep{Daskalaki2005,Luebbecke2008} to the reader, for comparison.

Multiphase exploitation of multiple objective-/value-restricted submodels
 in some aspects resembles very large neighbourhood search
 \citep{Ahuja2002} and adaptive large neighbourhood search
 \citep{Pisinger2006}: 
We use a number of different subproblems, which can be very large and slow to solve.
Multiple and/or large neighbourhoods have recently been very popular in local search solvers for timetabling applications.
\citet{DiGaspero2003} and \citet{Mueller2008}, for instance, use mutiple polynomial-time
 searchable neighbourhoods for Udine Course Timetabling. 
\citet{Abdullah2006JORS,Abdullah2007} and \citet{Meyers2006} have proposed large
 neighbourhoods for timetabling, without using integer programming to search the them.
\citet{Daskalaki2005} searches large neighbourhoods using integer programming.
\citet{Avella2007} study high-school timetabling problem and do a form of 
 ``value-restricted diving'' based on fixing the schedule for all but two teachers.
Depending on the particular problem and implementation, however, the ``surface'' 
 component in our approach tends to produce only few progressively improving solutions,
 and good lower bounds valid for the full problem,  
 which is often not the case in general large neighbourhood search. 
In contrast to large neighbourhood search, the ``dives'' also do not have to be
 explored at the time of their discovery or even in the order of discovery, or even on the same machine.
If it is possible to produce a number of progressively improving solutions
 at the surface, it indeed seems natural to start diving in neighbourhoods
 from the best solution found at the surface, with the hope that the
 solutions found within the dive can be used to cut off exploration of
 other dives once their lower bounds are known.  
Very large neighbourhood search nevertheless remains the metaheuristic approach 
 closest to the presented one.

There are also some similarities with the concept of ``ruin-and-recreate'' \citep{Schrimpf2002,IBM2002}.
As \cite{Lin1965} pointed out, 
 the key decision is which elements to preserve and which to relax,
 when the solution is passed from the surface to a dive.
This might be termed an option to ruin part of the solution before diving. 
However, it should be emphasised that the overall algorithm is very different
 from ``ruin-and-recreate''.
In particular there is no overall loop in which solutions are recycled.
A distinguishing feature or our approach is that the flow of solutions, as given in Figure~\ref{fig:info-flow},
 is different from the usual cycle of solutions.
More specifically, in Figure~\ref{fig:info-flow}, there is no arrow returning from the 
 dive to the surface level. 
The surface merely continues to produce its next solution, without reusing
 the result of the dive directly. 
If the present-best solution is found in a dive, its value only provides an 
 upper bound (cut-off limit) for further dives. 
This strategy seems much more compatible with the architecture of integer programming solvers.
 
There are also connections to column generation and dimensionality reduction
 in multiobjective optimisation. 
The heuristic surface-dive decomposition can be likened to column generation 
 and other exact decompositions studied in integer programming \citep{Ralphs2006MP}.
The ``surface'' and ``dive'' components could also be thought of as an biobjective integer program.
Although there has been recently some progress in the study of biobjective integer programs \citep{Ralphs2006AOR},
 they do not seem to yield direct performance improvements. 
It would also be most interesting to study the links to dimensionality reduction
 \citep{Tenenbaum2000}.
These connections, however, are largely yet to be established, which might take
 some time: As has been pointed out by \citet{Gandibleux2005}, the progress in understanding 
 the true multiobjective nature of many optimisation problems and related
 solution methods remains painfully slow.

%% file: results.tex
\begin{landscape}
\begin{table}
\centering
\renewcommand{\arraystretch}{1.3}
\caption{Linear programming relaxations for instances of the Udine Course Timetabling problem: 
dimensions of matrices after all default automatic reductions in CPLEX 11, 
root relaxation time using CPLEX 11 Dual Simplex LP solver,
and the remaining ``elapsed time'', after which CPLEX 11 is still at the root node of the monolithic formulation. 
For dives, the measurements are for the dive executed first in the given run.  
}
\label{tab:lp-relaxations}
\vskip 6mm
\input{tables/lp-relaxations2}
\end{table}
\end{landscape}

\begin{figure}[!h]
\centering
\renewcommand{\arraystretch}{1.1}
\caption{Our results compared against the results of \citet{Luebbecke2008} and \citet{Mueller2008}.
In all cases, CPLEX 11 is used with default parameters,
no custom cuts, and CPU time normalised using the benchmark of \citet{DiGaspero2007TR}.
Within 1 CPU unit, we use \textbf{Surface} and \textbf{PeriodFixed}.
Within 10 or 40 CPU units, we use \textbf{Surface2} with both \textbf{PeriodFixed} and \textbf{DayFixed} dives.
}
\vskip 6mm
{\small 
\input{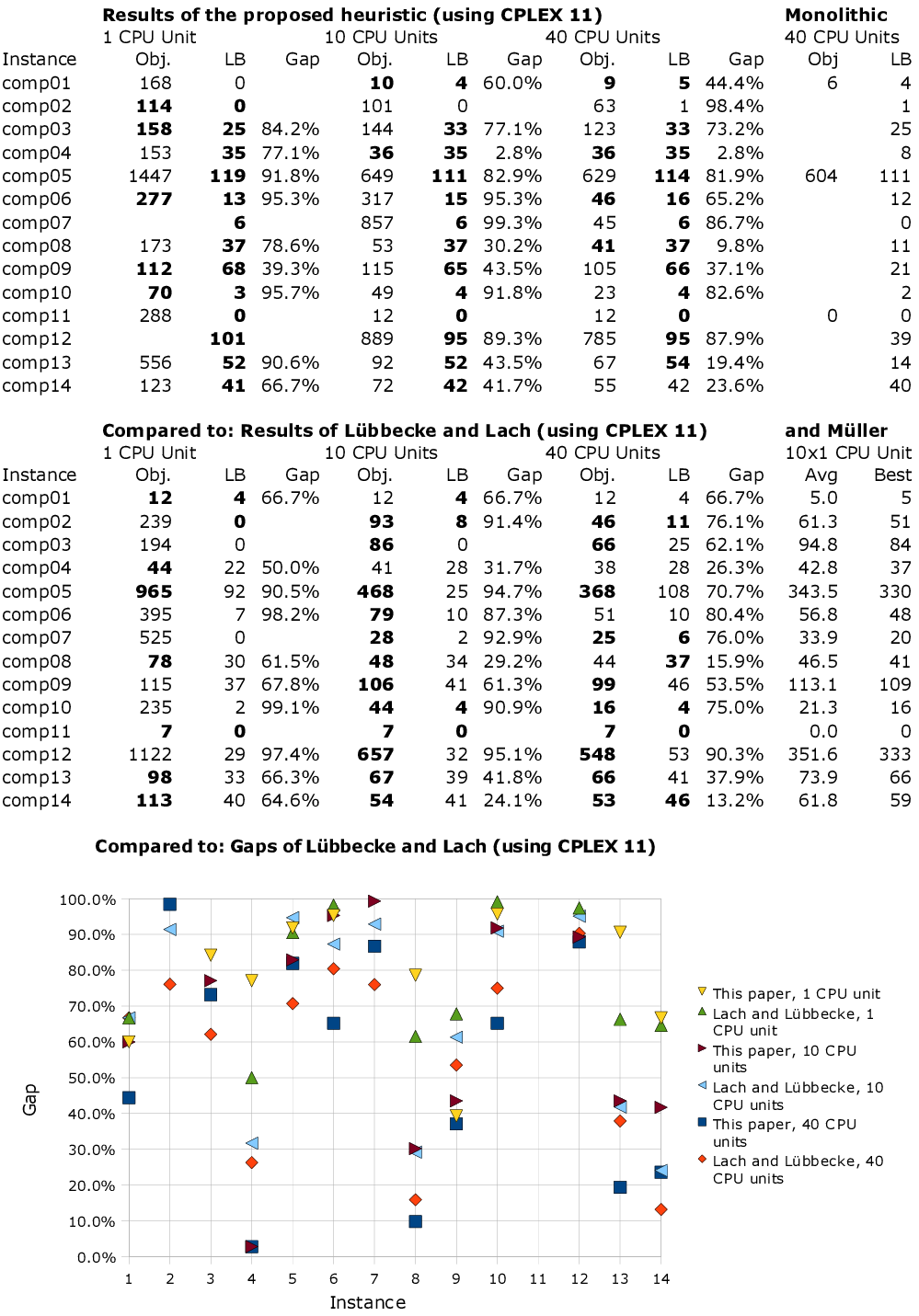}
}
\label{fig:results}
\end{figure}

\section{Computational Experience}
\label{sec:results}

The proposed heuristic has been implemented using ILOG Concert libraries in C++ \citep{cplex11apt}.
ILOG CPLEX 11 is used both at \textbf{Surface} and in all dives.
Some tests have also been carried out using
 ZIB SCIP, the present-best freely available integer programming solver developed 
 at Konrad-Zuse-Zentrum f{\"u}r Informationstechnik in Berlin \citep{Achterberg2004}; 
 their results are available on request from the authors.  
The implementation has been evaluated on a single processor of a desktop PC equipped with two 
  Intel Pentium 4 processors clocked at 3.20 GHz and with 4 GB of RAM, running Linux. 
To allow easier comparison with the results of \citet{Luebbecke2008} and entrants of 
 the International Timetabling Competition 2007,
 time measurements have been normalised using the benchmark\footnote{\url{http://www.cs.qub.ac.uk/itc2007/benchmarking/}} 
  by \citet{DiGaspero2007TR}.
One CPU unit in Figure~\ref{fig:results} corresponds to 780 seconds.
Within the limit of 1 CPU unit, we spend 600 seconds solving \textbf{Surface} and then 180 seconds in a single \textbf{PeriodFixed} dive.
Within the limit of 10 CPU units, we spend 3420 seconds solving \textbf{Surface2}, 
  then 180 seconds in a single \textbf{PeriodFixed} dive,
  and then 3600 seconds in a single \textbf{DayFixed} dive.
Within the limit of 40 CPU units, we spend 3 hours solving \textbf{Surface2},
  limited time in a number of \textbf{PeriodFixed} dives,
  and 5 hours in a single \textbf{DayFixed} dive.
Full source code, logs, and solution files are available from the contact author\footnote{\url{http://cs.nott.ac.uk/~jxm/} (Jakub Mare{\v c}ek)}.  

Overall, the solver can produce feasible solutions and corresponding lower bounds 
 for all instances from the International Timetabling Competition 2007 within one or two CPU units.
(It is only instances comp07 and comp12 that take two CPU units.)    
Indeed, if some terms of the objective are ignored at \textbf{Surface}, then provided the relative weights 
 of the other terms are not changed, the lower bound of the objective-restricted 
 problem is also a valid lower-bound for the complete minimisation problem.
However, within one CPU unit, alloted to solvers in the International 
  Timetabling Competition 2007, the progress is limited, mostly due to the
  sheer size of the LP instances involved, as outlined in Table~\ref{tab:lp-relaxations}.
The gap\footnote{The gap is calculated using the formula of ILOG, i.e. $100 (1 - LB/UB)$, where $LB$ is lower bound and $UB$ is cost of the best integer feasible solution.}
  often remains as large as 95 \%, if any solutions are found at all.
The solver nevertheless continues to make good progress, as suggested in Figure~\ref{fig:results},
  and within ten and 40 CPU units produces results with gaps comparable to those of \citet{Luebbecke2008}.
Compared to the results of \citet{Mueller2008}, the winner of Track 3 in the
  International Timetabling Competition 2007, the quality of the solutions obtained
  is only mediocre, no matter whether we consider the choice of the best result in 
  ten separate runs of one CPU unit each equivalent to ten CPU units, as suggested by
  \citet{Luebbecke2008}, or not.
Except for the instance comp04, comp06, and comp08, we have not been able to improve upper bounds obtained by
  \citet{Mueller2008} within 10 runs of 1 CPU unit within a single 40 CPU unit run.
On comp04, however, we have closed the absolute gap to 1 (2.8\%), a feat which seems
  rather difficult to achieve using monolithic formulations.
The possibility to perform a heuristic search whilst obtaining some information about its quality,
  perhaps together with the ease of implementation, hence represents two of the few,
 and perhaps the only two significant advantages of the proposed approach over standard 
 local search approaches.
Nevertheless, it seems that, given this performance, such an approach could be well usable in practice, for
  example, in tightly constrained instances, where solutions delivered by local search heuristics are
  not good enough to accept, without proofs of their near-optimality.
  
\subsection{Remarks on the Instances of Linear Programming}

In this connection, it is also interesting to notice the properties of the 
 linear programming relaxations involved, summarised in Table~\ref{tab:lp-relaxations}.
Generally, instances both at \textbf{Surface} and in all dives are considerably
 smaller than instances of \textbf{Monolithic},
 and there seem to be a number of instances (comp06, comp07), where the linear programming solver
 based on barrier methods is, by orders of magnitude, faster.   
On difficult instances, such as comp06, the run time of the linear programming solver
  for \textbf{Surface2} can be as much as four times as large as for \textbf{Surface},
  and similarly many iterations can be performed per second later on.
We do not have a good explanation for this behaviour. 
Compared to either \textbf{Surface} or \textbf{Surface2}, instances encountered in dives are very different. 
They are structurally more similar to weighted matching problems or multiple knapsack problems,
 often used to test integer programming solvers.
\textbf{PeriodFixed} dives can often be solved to optimality within a couple of 
 minutes on instances from the International Timetabling Competition 2007.
\textbf{DayFixed} dives often produce considerably better solutions within ten minutes on
 easier instances, but often do not provide any feasible solutions 
 within the first 30 minutes of the search on more difficult instances.
Improving the numerical behaviour of linear programming solvers on the instances
 could perhaps lead to considerable improvements in performance.
 
\section{Conclusions}
\label{sec:conc}

We have proposed a hybrid metaheuristic with an application in university course timetabling  
 with multiple soft constraints.
In addition to performing well on the particular problem of Udine Course Timetabling,
 this metaheuristic seems to provide a good starting point for the development of
 solvers for a number of similar problems with soft constraints with only minor modifications,
 including most timetabling problems with pattern penalisation. 
The hybridisation we study might also be of more general interest.
It is based on a mix of formulations of different sub-problems of the full problem,
 formulated in integer programming. 
The basic lesson learned is that the advantage of such hybridisation arises when 
 the sub-problems are selected so as to have good properties that can help the 
 integer programming solver. 
Specifically, the non-automatic reformulation of sub-problems is essential 
 in order to gain full advantage.
This is reminiscent of the dictum in local search, that
 the selection of neighbourhoods generally ought to be influenced by
 the issue of how effectively they can be implemented.
Overall, this approach seems to present a promising direction in the development 
 of hybrid metaheuristics.  

We believe that realistic problems often have many different terms in
the objectives and that often these objective terms can be quite
awkward to encode in integer programming, requiring many extra
variables. In this work on timetabling, we have seen an example of
such 'messy' problems.  We have seen that many different formulations,
relaxations and abstractions can play a role. Furthermore, with 
control of their usage, results can be improved significantly with
respect to direct use of the initial direct encoding. However, the
number of possible ways to create relaxations and abstractions and
link them together increases rapidly, and managing this can quickly
become unyieldy. Hence, a long-term future goal is to provide
intelligent decision support to help the user keep track of the
possibilties and semi-automatically discover good combinations. Our
belief is that such a system would greatly help integer programming
methods to achieve good results with less time and effort spent in
`tuning' the combinations of encodings.

\section*{Acknowledgment}
The authors are grateful
to Andrea Schaerf and Luca Di Gaspero, who kindly maintain the Udine Course 
Timetabling Problem, including all data sets.
Andrew Parkes has been supported by the UK Engineering and Physical Sciences
Research Council under grant GR/T26115/01.
Hana Rudov{\' a} has been supported by project MSM0021622419 of Ministerstvo 
{\v s}kolstv{\' i}, ml{\' a}de{\v z}e a t{\v e}lov{\' y}chovy and 
by the Grant Agency of the Czech Republic under project 201/07/0205.  
Last but not least, the authors are grateful to two anonymous referees for
  their insightful suggestions.


%% file: tables/lp-relaxations2.tex
\begin{tabular}{l|cr|cr|cr|cr}
Instance & \multicolumn{2}{c}{Reduced LP (\textbf{Monolithic})} & 
                                                \multicolumn{2}{c}{Reduced LP (\textbf{Surface})} & 
                                                                                 \multicolumn{2}{c}{Reduced LP (\textbf{PeriodFixed})} & 
                                                                                                                     \multicolumn{2}{c}{Reduced LP (\textbf{DayFixed})} \\ \hline
comp01 & $ 8593 \times 6696 $ & 29.01 s &     $ 3807 \times 2341 $ & 2.01 s &    $ 1484 \times 1128 $ & 0.14 s &     $ 3984 \times 4017 $ & 5.33 s \\ 
comp02 & $ 34438 \times 29559 $ & 1058.47 s & $ 8629 \times 5364 $ & 9.11 s &    $ 6605 \times 5840 $ & 2.80 s &     $ 8089 \times 10516 $ & 16.76 s \\ 
comp03 & $ 30985 \times 27240 $ & 493.43 s &  $ 7838 \times 4966 $ & 5.22 s &    $ 5825 \times 5136 $ & 1.92 s &     $ 7516 \times 10230 $ & 19.00 s \\ 
comp04 & $ 37570 \times 33195 $ & 590.15 s &  $ 8097 \times 5106 $ & 1.51 s &    $ 7367 \times 6570 $ & 2.29 s &     $ 8991 \times 14019 $ & 19.53 s \\ 
comp05 & $ 19795 \times 16652 $ & 109.08 s &  $ 9369 \times 6946 $ & 9.16 s &    $ 2282 \times 1827 $ & 0.04 s &     $ 5526 \times 5474 $ & 2.85 s \\ 
comp06 & $ 49360 \times 43485 $ & 1729.20 s & $ 10689 \times 6645 $ & 16.63 s &  $ 9361 \times 8442 $ & 9.26 s &     $13096 \times 19977 $ & 78.90 s \\ 
comp07 & $ 67901 \times 60001 $ & 6489.69 s & $ 13491 \times 8108 $ & 45.05 s &  $ 12365 \times 11300 $ & 13.72 s &  $ 16974 \times 24940 $ & 157.06 s \\ 
comp08 & $ 39935 \times 35224 $ & 899.81 s &  $ 8643 \times 5434 $ & 2.63 s &    $ 8240 \times 7380 $ & 6.02 s &     $11316 \times 17657 $ & 42.66 s \\ 
comp09 & $ 36382 \times 31951 $ & 653.75 s &  $ 8290 \times 5333 $ & 2.19 s &    $ 7195 \times 6390 $ & 3.75 s &     $8717 \times 11881 $ & 17.58 s \\ 
comp10 & $ 51839 \times 45759 $ & 2087.70 s & $ 11244 \times 6850 $ & 21.58 s &  $ 9628 \times 8712 $ & 8.97 s &     $ 12388 \times 20637 $ & 83.90 s \\ 
comp11 & $ 11747 \times 8448 $ & 28.86 s &    $ 5699 \times 3334 $ & 2.93 s &    $ 1367 \times 960 $ & 0.06 s &      $ 5352 \times 5896 $ & 6.56 s \\ 
comp12 & $ 32893 \times 27504 $ & 480.57 s &  $ 12570 \times 8724 $ & 27.25 s &  $ 3990 \times 3355 $ & 0.30 s &     $ 8114 \times 8342 $ & 10.84 s \\ 
comp13 & $ 39610 \times 35188 $ & 771.64 s &  $ 8320 \times 5329 $ & 2.33 s &    $ 8275 \times 7410 $ & 4.33 s &     $ 10599 \times 16795 $ & 33.44 s \\ 
comp14 & $ 37785 \times 32828 $ & 866.63 s &  $ 8685 \times 5333 $ & 8.10 s &    $ 6905 \times 6120 $ & 2.23 s &     $ 8151 \times 12674 $ & 13.19 s  
\end{tabular}

%% file: tables/results.tex
\begin{tabular}{l|lllllllll|ll} \hline
	& \multicolumn{9}{c}{Results of the proposed heuristic (using CPLEX 11)} & \multicolumn{2}{c}{Monolithic} \\
	& \multicolumn{3}{c}{1 CPU Unit}	& \multicolumn{3}{c}{10 CPU Units}	& \multicolumn{3}{c}{40 CPU Units} & \multicolumn{2}{c}{40 CPU Units} \\ 
Instance	& Obj.	& LB	& Gap	& Obj.	& LB	& Gap	& Obj.	& LB	& Gap	& Obj	& LB \\  \hline
comp01	& 168	& 0	& 	& 10	& 4	& 60.0\%	& 9	& 5	& 44.4\%	& 6	& 4 \\
comp02	& 114	& 0	& 	& 101	& 0	& 	& 63	& 1	& 98.4\%	& 	& 1 \\
comp03	& 158	& 25	& 84.2\%	& 144	& 33	& 77.1\%	& 123	& 33	& 73.2\%	& 	& 25 \\
comp04	& 153	& 35	& 77.1\%	& 36	& 35	& 2.8\%	& 36	& 35	& 2.8\%	& 	& 8 \\
comp05	& 1447	& 119	& 91.8\%	& 649	& 111	& 82.9\%	& 629	& 114	& 81.9\%	& 604	& 111 \\
comp06	& 277	& 13	& 95.3\%	& 317	& 15	& 95.3\%	& 46	& 16	& 65.2\%	& 	& 12 \\
comp07	& 	& 6	& 	& 857	& 6	& 99.3\%	& 45	& 6	& 86.7\%	& 	& 0 \\
comp08	& 173	& 37	& 78.6\%	& 53	& 37	& 30.2\%	& 41	& 37	& 9.8\%	& 	& 11 \\
comp09	& 112	& 68	& 39.3\%	& 115	& 65	& 43.5\%	& 105	& 66	& 37.1\%	& 	& 21 \\
comp10	& 70	& 3	& 95.7\%	& 49	& 4	& 91.8\%	& 23	& 4	& 82.6\%	& 	& 2 \\
comp11	& 288	& 0	& 	& 12	& 0	& 	& 12	& 0	& 	& 0	& 0 \\
comp12	& 	& 101	& 	& 889	& 95	& 89.3\%	& 785	& 95	& 87.9\%	& 	& 39 \\
comp13	& 556	& 52	& 90.6\%	& 92	& 52	& 43.5\%	& 67	& 54	& 19.4\%	& 	& 14 \\
comp14	& 123	& 41	& 66.7\%	& 72	& 42	& 41.7\%	& 55	& 42	& 23.6\%	& 	& 40 \\  \hline
	& \multicolumn{9}{c}{Compared to: Results of L{\" u}bbecke and Lach (using CPLEX 11)} & \multicolumn{2}{c}{and M{\" u}ller} \\ 
	& \multicolumn{3}{c}{1 CPU Unit}	& \multicolumn{3}{c}{10 CPU Units}	& \multicolumn{3}{c}{40 CPU Units}	& \multicolumn{2}{c}{10x1 CPU Unit} \\  
Instance	& Obj.	& LB	& Gap	& Obj.	& LB	& Gap	& Obj.	& LB	& Gap	& Avg	& Best \\  \hline
comp01	& 12	& 4	& 66.7\%	& 12	& 4	& 66.7\%	& 12	& 4	& 66.7\%	& 5.0	& 5 \\
comp02	& 239	& 0	& 	& 93	& 8	& 91.4\%	& 46	& 11	& 76.1\%	& 61.3	& 51 \\
comp03	& 194	& 0	& 	& 86	& 0	& 	& 66	& 25	& 62.1\%	& 94.8	& 84 \\
comp04	& 44	& 22	& 50.0\%	& 41	& 28	& 31.7\%	& 38	& 28	& 26.3\%	& 42.8	& 37 \\
comp05	& 965	& 92	& 90.5\%	& 468	& 25	& 94.7\%	& 368	& 108	& 70.7\%	& 343.5	& 330 \\
comp06	& 395	& 7	& 98.2\%	& 79	& 10	& 87.3\%	& 51	& 10	& 80.4\%	& 56.8	& 48 \\
comp07	& 525	& 0	& 	& 28	& 2	& 92.9\%	& 25	& 6	& 76.0\%	& 33.9	& 20 \\
comp08	& 78	& 30	& 61.5\%	& 48	& 34	& 29.2\%	& 44	& 37	& 15.9\%	& 46.5	& 41 \\
comp09	& 115	& 37	& 67.8\%	& 106	& 41	& 61.3\%	& 99	& 46	& 53.5\%	& 113.1	& 109 \\
comp10	& 235	& 2	& 99.1\%	& 44	& 4	& 90.9\%	& 16	& 4	& 75.0\%	& 21.3	& 16 \\
comp11	& 7	& 0	& 	& 7	& 0	& 	& 7	& 0	& 	& 0.0	& 0 \\
comp12	& 1122	& 29	& 97.4\%	& 657	& 32	& 95.1\%	& 548	& 53	& 90.3\%	& 351.6	& 333 \\
comp13	& 98	& 33	& 66.3\%	& 67	& 39	& 41.8\%	& 66	& 41	& 37.9\%	& 73.9	& 66 \\
comp14	& 113	& 40	& 64.6\%	& 54	& 41	& 24.1\%	& 53	& 46	& 13.2\%	& 61.8	& 59  \\ \hline
\end{tabular}

%% file: diving.bbl
\begin{thebibliography}{76}
\expandafter\ifx\csname natexlab\endcsname\relax\def\natexlab#1{#1}\fi
\expandafter\ifx\csname url\endcsname\relax
  \def\url#1{\texttt{#1}}\fi
\expandafter\ifx\csname urlprefix\endcsname\relax\def\urlprefix{URL }\fi

\bibitem[{Abdullah et~al.(2007)Abdullah, Ahmadi, Burke, and
  Dror}]{Abdullah2007}
Abdullah, S., Ahmadi, S., Burke, E.~K., Dror, M., 2007. Investigating
  {A}huja--{O}rlin's large neighbourhood search approach for examination
  timetabling. OR Spectrum 29~(2), 351--372.

\bibitem[{Abdullah et~al.(2006)Abdullah, Ahmadi, Burke, Dror, and
  McCollum}]{Abdullah2006JORS}
Abdullah, S., Ahmadi, S., Burke, E.~K., Dror, M., McCollum, B., 2006. A
  tabu-based large neighbourhood search methodology for the capacitated
  examination timetabling problem. J. Oper. Res. Soc. 58, 1494–--1502.

\bibitem[{Achterberg(2004)}]{Achterberg2004}
Achterberg, T., 2004. {SCIP} -- {A} framework to integrate constraint and mixed
  integer programming. ZIB Technical Report TR-04-19.

\bibitem[{Ahuja et~al.(2002)Ahuja, Ergun, Orlin, and Punnen}]{Ahuja2002}
Ahuja, R.~K., Ergun, {\" O}., Orlin, J.~B., Punnen, A.~P., 2002. A survey of
  very large-scale neighborhood search techniques. Discrete Appl. Math.
  123~(1-3), 75--102.

\bibitem[{Al-Yakoob and Sherali(2007)}]{AlYakoob2007}
Al-Yakoob, S.~M., Sherali, H.~D., 2007. A mixed-integer programming approach to
  a class timetabling problem: {A} case study with gender policies and traffic
  considerations. European J. Oper. Res., In press.

\bibitem[{Avella et~al.(2007)Avella, D'Auria, Salerno, and
  Vasil'Ev}]{Avella2007}
Avella, P., D'Auria, B., Salerno, S., Vasil'Ev, I., 2007. A computational study
  of local search algorithms for italian high-school timetabling. Journal of
  Heuristics 13~(6), 543--556.

\bibitem[{Avella and Vasil'ev(2005)}]{Avella2005}
Avella, P., Vasil'ev, I., 2005. A computational study of a cutting plane
  algorithm for university course timetabling. J. Scheduling 8~(6), 497--514.

\bibitem[{Bardadym(1996)}]{Bardadym1996}
Bardadym, V.~A., 1996. Computer-aided school and university timetabling: The
  new wave. In: Burke, E.~K., Ross, P. (Eds.), Practice and Theory of Automated
  Timetabling. Vol. 1153 of Lecture Notes in Computer Science. Springer,
  Berlin, pp. 22--45.

\bibitem[{Barnhart et~al.(1993)Barnhart, Johnson, Nemhauser, Sigismondi, and
  Vance}]{Nemhauser1993}
Barnhart, C., Johnson, E.~L., Nemhauser, G.~L., Sigismondi, G., Vance, P.,
  1993. Formulating a mixed integer programming problem to improve solvability.
  Oper. Res. 41~(6), 1013--1019.

\bibitem[{Berthold(2007{\natexlab{a}})}]{Berthold2007}
Berthold, T., 2007{\natexlab{a}}. Heuristics of the
  branch-cut-and-price-framework {SCIP}. Tech. rep., Berlin, zIB Technical
  Report TR-07-30.

\bibitem[{Berthold(2007{\natexlab{b}})}]{Berthold2008}
Berthold, T., 2007{\natexlab{b}}. {RENS} -- {R}elaxation enforced neighborhood
  search. Tech. rep., Berlin, zIB Technical Report TR-07-28.

\bibitem[{Beyrouthy et~al.(2007)Beyrouthy, Burke, Landa-Silva, McCollum,
  McMullan, and Parkes}]{Beyrouthy2007}
Beyrouthy, C., Burke, E.~K., Landa-Silva, D., McCollum, B., McMullan, P.,
  Parkes, A.~J., 2007. Towards improving the utilisation of university teaching
  space. J. Oper. Res. Soc., to appear.

\bibitem[{Bixby et~al.(2004)Bixby, Fenelon, Gu, Rothberg, and
  Wunderling}]{Bixby2004}
Bixby, R.~E., Fenelon, M., Gu, Z., Rothberg, E., Wunderling, R., 2004.
  Mixed-integer programming: A progress report. In: Gr\"{o}tschel, M. (Ed.),
  The Sharpest Cut: {T}he Impact of {M}anfred {P}adberg and His Work. SIAM,
  Philadelphia, PA.

\bibitem[{Blum and Roli(2003)}]{Blum2003}
Blum, C., Roli, A., 2003. Metaheuristics in combinatorial optimization:
  Overview and conceptual comparison. ACM Comput. Surv. 35~(3), 268--308.

\bibitem[{Bodlaender and Fomin(2005)}]{Bodlaender2005}
Bodlaender, H.~L., Fomin, F.~V., 2005. Equitable colorings of bounded treewidth
  graphs. Theor. Comput. Sci. 349~(1), 22--30.

\bibitem[{Burke and Newall(1999)}]{Burke1999a}
Burke, E., Newall, J., Apr 1999. A multi-stage evolutionary algorithm for the
  timetable problem. IEEE Transactions on Evolutionary Computation 13~(1),
  63--74.

\bibitem[{Burke and Causmaecker(2003)}]{patat2002}
Burke, E.~K., Causmaecker, P.~D. (Eds.), 2003. Practice and Theory of Automated
  Timetabling, 4th International Conference, {PATAT} 2002, {G}ent, {B}elgium,
  {A}ugust 21-23, 2002, {S}elected Revised Papers. Vol. 2740 of Lecture Notes
  in Computer Science. Springer, Berlin.

\bibitem[{Burke et~al.(2004)Burke, de~Werra, and Kingston}]{Burke2004}
Burke, E.~K., de~Werra, D., Kingston, J.~H., 2004. Applications to timetabling.
  In: Gross, J.~L., Yellen, J. (Eds.), Handbook of Graph Theory. CRC, London,
  UK, pp. 445--474.

\bibitem[{Burke and Gendreau(2008)}]{patat2008}
Burke, E.~K., Gendreau, M. (Eds.), 2008. Practice and Theory of Automated
  Timetabling, 7th International Conference, {PATAT} 2008, {M}ontr{\' e}al,
  {C}anada, {A}ugust 19--22, 2008, {P}roceedings. Universit{\' e} de Montr{\'
  e}al, Montr{\' e}al, Canada.

\bibitem[{Burke et~al.(1997)Burke, Jackson, Kingston, and Weare}]{Burke1997}
Burke, E.~K., Jackson, K., Kingston, J.~H., Weare, R.~F., 1997. Automated
  university timetabling: {T}he state of the art. Comput. J. 40~(9), 565--571.

\bibitem[{Burke et~al.(2007)Burke, Mare{\v c}ek, Parkes, and Rudov{\'
  a}}]{Marecek2007TR}
Burke, E.~K., Mare{\v c}ek, J., Parkes, A.~J., Rudov{\' a}, H., 2007. On a
  clique-based integer programming formulation of vertex colouring with
  applications in course timetabling. Tech. Rep. {NOTTCS-TR-2007-10}, The
  University of Nottingham, Nottingham, also available at
  http://arxiv.org/abs/0710.3603.

\bibitem[{Burke et~al.(2008{\natexlab{a}})Burke, Mare{\v c}ek, Parkes, and
  Rudov{\' a}}]{Marecek2008TR}
Burke, E.~K., Mare{\v c}ek, J., Parkes, A.~J., Rudov{\' a}, H.,
  2008{\natexlab{a}}. A branch-and-cut procedure for {U}dine {C}ourse
  {T}imetabling. In:  \cite{patat2008}, also appeared as NOTTCS-TR-2008-1.

\bibitem[{Burke et~al.(2008{\natexlab{b}})Burke, Mare{\v c}ek, Parkes, and
  Rudov{\' a}}]{Marecek2007GOR}
Burke, E.~K., Mare{\v c}ek, J., Parkes, A.~J., Rudov{\' a}, H.,
  2008{\natexlab{b}}. Penalising patterns in timetables: Novel integer
  programming formulations. In: Nickel, S., Kalcsics, J. (Eds.), Operations
  Research Proceedings 2007. Springer, Berlin, pp. 409--414.

\bibitem[{Burke and Petrovic(2002)}]{Petrovic2002}
Burke, E.~K., Petrovic, S., 2002. Recent research directions in automated
  timetabling. European J. Oper. Res. 140~(2), 266--280.

\bibitem[{Burke and Rudov{\' a}(2007)}]{patat2007}
Burke, E.~K., Rudov{\' a}, H. (Eds.), 2007. Practice and Theory of Automated
  Timetabling, 6th International Conference, {PATAT} 2006, {B}rno, The {C}zech
  {R}epublic, {A}ugust 30 -- {S}eptember 1, 2006, {S}elected {R}evised
  {P}apers. Vol. 3867 of Lecture Notes in Computer Science. Springer, Berlin.

\bibitem[{Cal\'{e}gari et~al.(1999)Cal\'{e}gari, Coray, Hertz, Kobler, and
  Kuonen}]{Celegari1999}
Cal\'{e}gari, P., Coray, G., Hertz, A., Kobler, D., Kuonen, P., 1999. A
  taxonomy of evolutionary algorithms in combinatorial optimization. J.
  Heuristics 5~(2), 145--158.

\bibitem[{Carter(1989)}]{Carter1989}
Carter, M.~W., 1989. A {L}agrangian relaxation approach to the classroom
  assignment problem. INFORMS J. Comput. 27, 230--245.

\bibitem[{Carter and Laporte(1998)}]{Carter1997}
Carter, M.~W., Laporte, G., 1998. Recent developments in practical course
  timetabling. In: Burke, E.~K., Carter, M.~W. (Eds.), Practice and Theory of
  Automated Timetabling. Vol. 1408 of Lecture Notes in Computer Science.
  Springer, Berlin, pp. 3--19.

\bibitem[{Chabrier(2006)}]{Chabrier2006}
Chabrier, A., 2006. Vehicle routing problem with elementary shortest path based
  column generation. Comput. Oper. Res. 33~(10), 2972--2990.

\bibitem[{Danna et~al.(2005)Danna, Rothberg, and Pape}]{Danna2005}
Danna, E., Rothberg, E., Pape, C.~L., 2005. Exploring relaxation induced
  neighborhoods to improve {MIP} solutions. Math. Program. 102~(1), 71--90.

\bibitem[{Daskalaki et~al.(2004)Daskalaki, Birbas, and Housos}]{Daskalaki2004}
Daskalaki, S., Birbas, T., Housos, E., 2004. An integer programming formulation
  for a case study in university timetabling. European J. Oper. Res. 153,
  117--135.

\bibitem[{Daskalaki et~al.(2005)Daskalaki, Birbas, and Housos}]{Daskalaki2005}
Daskalaki, S., Birbas, T., Housos, E., 2005. Efficient solutions for a
  university timetabling problem through integer programming. European J. Oper.
  Res. 160, 106--120.

\bibitem[{{Di~Gaspero} et~al.(2007){Di~Gaspero}, McCollum, , and
  Schaerf}]{DiGaspero2007TR}
{Di~Gaspero}, L., McCollum, B., , Schaerf, A., 2007. The second international
  timetabling competition ({ITC}-2007): Curriculum-based course timetabling
  ({T}rack 3). Tech. Rep. QUB/IEEE 2007/08/01, University of Udine DIEGM,
  Udine, Italy.

\bibitem[{{Di~Gaspero} and Schaerf(2003)}]{DiGaspero2003}
{Di~Gaspero}, L., Schaerf, A., 2003. Multi neighborhood local search with
  application to the course timetabling problem. In:  \cite{patat2002}, pp.
  262--275.

\bibitem[{{Di~Gaspero} and Schaerf(2006)}]{DiGaspero2006}
{Di~Gaspero}, L., Schaerf, A., 2006. Neighborhood portfolio approach for local
  search applied to timetabling problems. J. Math. Model. Algorithms 5~(1),
  65--89.

\bibitem[{Dimopoulou and Miliotis(2004)}]{Dimopoulou2004}
Dimopoulou, M., Miliotis, P., 2004. An automated university course timetabling
  system developed in a distributed environment: {A} case study. European J.
  Oper. Res. 153, 136--147.

\bibitem[{Dueck et~al.(2002)Dueck, Maehler, Schneider, Schrimpf, and
  Stamm-Wilbrandt}]{IBM2002}
Dueck, G., Maehler, M., Schneider, J., Schrimpf, G., Stamm-Wilbrandt, H., 2002.
  Optimization with ruin recreate. US Patent 6418398.

\bibitem[{El-Abd and Kamel(2005)}]{ElAbd2005}
El-Abd, M., Kamel, M., 2005. A taxonomy of cooperative search algorithms. In:
  Hybrid Metaheuristics. Vol. 3636 of Lecture Notes in Computer Science.
  Springer, Berlin, pp. 32--41.

\bibitem[{Even et~al.(1976)Even, Itai, and Shamir}]{Even1976}
Even, S., Itai, A., Shamir, A., 1976. On the complexity of timetable and
  multicommodity flow problems. SIAM J. Computing 5~(4), 691--703.

\bibitem[{Fischetti et~al.(2005)Fischetti, Glover, and Lodi}]{Fischetti2005}
Fischetti, M., Glover, F., Lodi, A., 2005. The feasibility pump. Math. Program.
  104~(1), 91--104.

\bibitem[{Fischetti and Lodi(2003)}]{Fischetti2003}
Fischetti, M., Lodi, A., 2003. Local branching. Math. Program. 98~(1), 23--47.

\bibitem[{Forrest et~al.(2004)Forrest, de~la Nuez, and
  Lougee-Heimer}]{Forrest2004}
Forrest, J., de~la Nuez, D., Lougee-Heimer, R., 2004. {CLP} {U}ser guide. Tech.
  rep., Yorktown Heights, NY.

\bibitem[{Gandibleux and Ehrgott(2005)}]{Gandibleux2005}
Gandibleux, X., Ehrgott, M., 2005. 1984--2004 -- 20 years of multiobjective
  metaheuristics. {B}ut what about the solution of combinatorial problems with
  multiple objectives? In: Coello, C. A.~C., Aguirre, A.~H., Zitzler, E.
  (Eds.), Evolutionary Multi-Criterion Optimization. Vol. 3410 of Lecture Notes
  in Computer Science. Springer, Berlin, pp. 33--46.

\bibitem[{Garey and Johnson(1979)}]{Garey1979}
Garey, M.~R., Johnson, D.~S., 1979. Computers and Intractability: {A} Guide to
  the Theory of NP-Completeness. W. H. Freeman \& Co., New York, NY.

\bibitem[{ILOG(2007)}]{cplex11apt}
ILOG, 2007. {ILOG CPLEX} {A}dvanced Programming Techniques. ILOG S. A., Incline
  Village, NV.

\bibitem[{Johnson et~al.(2000)Johnson, Nemhauser, and
  Savelsbergh}]{Nemhauser2000}
Johnson, E.~L., Nemhauser, G.~L., Savelsbergh, M.~W., 2000. Progress in linear
  programming-based algorithms for integer programming: An exposition. INFORMS
  J. Computing 12~(1), 2--23.

\bibitem[{Lach and L{\" u}bbecke(2008{\natexlab{a}})}]{Luebbecke2008}
Lach, G., L{\" u}bbecke, M.~E., 2008{\natexlab{a}}. Curriculum based course
  timetabling: Optimal solutions to the udine benchmark instances. In:
  \cite{patat2008}.

\bibitem[{Lach and L{\" u}bbecke(2008{\natexlab{b}})}]{Lach2008}
Lach, G., L{\" u}bbecke, M.~E., 2008{\natexlab{b}}. Optimal university course
  timetables and the partial transversal polytope. In: McGeoch, C.~C. (Ed.),
  WEA. Vol. 5038 of Lecture Notes in Computer Science. Springer, pp. 235--248.

\bibitem[{Lawrie(1969)}]{Lawrie1969}
Lawrie, N.~L., 1969. An integer linear programming model of a school
  timetabling problem. Comput. J. 12, 307--316.

\bibitem[{Leung et~al.(2004)Leung, Kelly, and Anderson}]{Leung2004}
Leung, J., Kelly, L., Anderson, J.~H., 2004. Handbook of Scheduling:
  {A}lgorithms, Models, and Performance Analysis. CRC, Boca Raton, FL.

\bibitem[{Lin(1965)}]{Lin1965}
Lin, S., 1965. Computer solutions to the travelling salesman problems. Bell
  Syst. Tech. J. 44~(10), 2245--2269.

\bibitem[{McCollum(2007)}]{McCollum2006}
McCollum, B., 2007. A perspective on bridging the gap between theory and
  practice in university timetabling. In:  \cite{patat2007}, pp. 3--23.

\bibitem[{Meyers and Orlin(2006)}]{Meyers2006}
Meyers, C., Orlin, J.~B., 2006. Very large-scale neighborhood search techniques
  in timetabling problems. In: Practice and Theory of Automated Timetabling.
  pp. 24--39.

\bibitem[{Mirhassani(2006)}]{Mirhassani2006a}
Mirhassani, S.~A., 2006. A computational approach to enhancing course
  timetabling with integer programming. Appl. Math. Comput. 175, 814--822.

\bibitem[{M{\" u}ller(2008)}]{Mueller2008}
M{\" u}ller, T., 2008. Itc2007 solver description: A hybrid approach. In:
  \cite{patat2008}.

\bibitem[{Murray et~al.(2007)Murray, M{\" u}ller, and Rudov{\' a}}]{Rudova2007}
Murray, K., M{\" u}ller, T., Rudov{\' a}, H., 2007. Modeling and solution of a
  complex university course timetabling problem. In:  \cite{patat2007}, pp.
  189--209.

\bibitem[{Nemhauser and Wolsey(1988)}]{NemhauserWolsey1988}
Nemhauser, G.~L., Wolsey, L.~A., 1988. Integer and combinatorial optimization.
  John Wiley \& Sons, New York, NY.

\bibitem[{Petrovic and Burke(2004)}]{Petrovic2004}
Petrovic, S., Burke, E.~K., 2004. University timetabling. In:
  \cite{Leung2004}, pp. 1001--1023.

\bibitem[{Pisinger and R{\o}pke(2007)}]{Pisinger2007}
Pisinger, D., R{\o}pke, S., 2007. A general heuristic for vehicle routing
  problems. Comput. Oper. Res. 34, 2403--2435.

\bibitem[{Puchinger and Raidl(2005)}]{Puchinger2005}
Puchinger, J., Raidl, G.~R., 2005. Combining metaheuristics and exact
  algorithms in combinatorial optimization: {A} survey and classification. In:
  Mira, J., {\'A}lvarez, J.~R. (Eds.), IWINAC (2). Vol. 3562 of Lecture Notes
  in Computer Science. Springer, Berlin, pp. 41--53.

\bibitem[{Qualizza and Serafini(2005)}]{Qualizza2004}
Qualizza, A., Serafini, P., 2005. A column generation scheme for faculty
  timetabling. In: Burke, E.~K., Trick, M.~A. (Eds.), Practice and Theory of
  Automated Timetabling. Vol. 3616 of Lecture Notes in Computer Science.
  Springer, Berlin, pp. 161--173.

\bibitem[{Raidl(2006)}]{Raidl2006}
Raidl, G.~R., 2006. A unified view on hybrid metaheuristics. In: Almeida, F.,
  Aguilera, M. J.~B., Blum, C., Moreno-Vega, J.~M., P{\'e}rez, M.~P., Roli, A.,
  Sampels, M. (Eds.), Hybrid Metaheuristics. Vol. 4030 of Lecture Notes in
  Computer Science. Springer, Berlin, pp. 1--12.

\bibitem[{Ralphs and Galati(2006)}]{Ralphs2006MP}
Ralphs, T.~K., Galati, M.~V., 2006. Decomposition and dynamic cut generation in
  integer linear programming. Math. Program. 106~(2), 261--285.

\bibitem[{Ralphs et~al.(2006)Ralphs, Saltzman, and Wiecek}]{Ralphs2006AOR}
Ralphs, T.~K., Saltzman, M.~J., Wiecek, M.~M., 2006. An improved algorithm for
  solving biobjective integer programs. Ann. Oper. Res. 147, 43--70.

\bibitem[{R{\o}pke and Pisinger(2006)}]{Pisinger2006}
R{\o}pke, S., Pisinger, D., 2006. An adaptive large neighborhood search
  heuristic for the pickup and delivery problem with time windows. Transport.
  Sci. 40, 455--472.

\bibitem[{Rudov{\' a} and Murray(2003)}]{Rudova2002}
Rudov{\' a}, H., Murray, K., 2003. University course timetabling with soft
  constraints. In:  \cite{patat2002}, pp. 310--328.

\bibitem[{Schaerf(1999)}]{Schaerf1999}
Schaerf, A., 1999. A survey of automated timetabling. Artificial Intelligence
  Rev. 13~(2), 87--127.

\bibitem[{Schimmelpfeng and Helber(2007)}]{Helber2007}
Schimmelpfeng, K., Helber, S., 2007. Application of a real-world
  university-course timetabling model solved by integer programming. OR
  Spectrum 29, 783--803.

\bibitem[{Schrimpf et~al.(2000)Schrimpf, Schneider, Stamm-Wilbrandt, and
  Dueck}]{Schrimpf2002}
Schrimpf, G., Schneider, J., Stamm-Wilbrandt, H., Dueck, G., 2000. Record
  breaking optimization results using the ruin and recreate principle. J.
  Comput. Phys. 159~(2), 139--171.

\bibitem[{Tenenbaum et~al.(2000)Tenenbaum, de~Silva, and
  Langford}]{Tenenbaum2000}
Tenenbaum, J.~B., de~Silva, V., Langford, J.~C., 2000. A global geometric
  framework for nonlinear dimensionality reduction. Science 290, 2319--2323.

\bibitem[{Trick(2005)}]{Trick2005}
Trick, M.~A., 2005. Formulations and reformulations in integer programming. In:
  Bart{\' a}k, R., Milano, M. (Eds.), Integration of {AI} and {OR} Techniques
  in Constraint Programming for Combinatorial Optimization Problems. Vol. 3524
  of Lecture Notes in Computer Science. Springer, Berlin, pp. 366--379.

\bibitem[{Tripathy(1984)}]{Tripathy1984}
Tripathy, A., 1984. School timetabling -- {A} case in large binary integer
  linear programming. Management Sci. 30, 1473--1489.

\bibitem[{Welsh and Powell(1967)}]{Welsh1967}
Welsh, D. J.~A., Powell, M.~B., 1967. An upper bound for the chromatic number
  of a graph and its application to timetabling problems. Computer J. 10~(1),
  85--86.

\bibitem[{Williams(1978)}]{Williams1978}
Williams, H.~P., 1978. The reformulation of two mixed integer programming
  problems. Math. Programming 14~(3), 325--331.

\bibitem[{Wunderling(1996)}]{Wunderling1996}
Wunderling, R., 1996. Paralleler und {O}bjektorientierter
  {S}implex-{A}lgorithmus. ZIB Technical Report TR-96-09.

\bibitem[{Zilberstein(1996)}]{Zilberstein1996}
Zilberstein, S., 1996. Using anytime algorithms in intelligent systems. AI
  Magazine 17~(3), 73--83.

\end{thebibliography}
